\documentclass[preprint,12pt]{elsarticle}

\usepackage{amssymb}
\usepackage{amsmath}

\usepackage{booktabs,caption}
\usepackage[flushleft]{threeparttable}

\usepackage{graphicx}
\usepackage{caption}
\usepackage{subcaption}

\usepackage{color,soul}

\usepackage[utf8]{inputenc} 
\usepackage[T1]{fontenc}

\usepackage[skip=2.5pt,font=scriptsize]{caption}

\setcitestyle{authoryear,open={(},close={)}} 

\makeatletter
\def\ps@pprintTitle{%
  \let\@oddhead\@empty
  \let\@evenhead\@empty
  \let\@oddfoot\@empty
  \let\@evenfoot\@oddfoot
}
\makeatother

\journal{Icarus}
\begin{document}
\begin{frontmatter}

\title{A pole-to-pole map of hydrocarbons in Saturn's upper stratosphere and mesosphere}

\author[LPL]{Zarah L. Brown}
\ead{zarahbrown@arizona.edu}
\author[LPL]{Tommi T. Koskinen}
\author[SSI]{Julianne I. Moses}
\author[LMD,LEISA]{Sandrine Guerlet}

\affiliation[LPL]{organization={Lunar and Planetary Laboratory, University of Arizona},
            addressline={1629 E. University Blvd.}, 
            city={Tucson},
            postcode={85721}, 
            state={Arizona},
            country={U.S.A.}}
\affiliation[SSI]{organization={Space Science Institute},
            addressline={4765 Walnut St., Suite B}, 
            city={Boulder},
            postcode={80301}, 
            state={Colorado},
            country={U.S.A.}}
\affiliation[LMD]{organization={Laboratoire de M\'et\'eorologie Dynamique (LMD/IPSL), Sorbonne Universit\'e, Centre National de la Recherche Scientifique, \'Ecole Polytechnique, \'Ecole Normale Sup\'erieure},
            addressline={4 Place Jussieu},
            postcode={75005},
            city={Paris},
            country={France}}
\affiliation[LEISA]{organization={LESIA, Observatoire de Paris, Universit\'e PSL, CNRS, Sorbonne Universit\'e, Universit\'e de Paris},
            addressline={5 place Jules Janssen},
            postcode={92195},
            city={Paris},
            country={France}}

\begin{abstract}
We analyze data from the final two years of the Cassini mission to retrieve the distributions of methane (CH$_4$), ethane (C$_2$H$_6$), acetylene (C$_2$H$_2$), ethylene (C$_2$H$_4$), and benzene (C$_6$H$_6$) in Saturn's upper stratosphere and mesosphere from stellar occultations observed by the Ultraviolet Imaging Spectrograph (UVIS), spanning latitudes from pole to pole. These observations represent the first two-dimensional snapshot of the photochemical production region with latitude and depth for these five light hydrocarbons around the northern summer solstice. To support this analysis, we combine temperature-pressure profiles retrieved from the UVIS occultations and limb scans observed by the Cassini Composite Infrared Spectrometer (CIRS) with the CH$_4$ profiles to provide atmospheric structure models for each occultation location that span the middle and upper atmosphere. We detect a strong meridional trend in the homopause pressure level that implies much weaker mixing at the poles than near the subsolar point. This is shown by the homopause pressure level, which ranges from $\sim$0.05 $\mu$bar around the subsolar point to $\sim$5 $\mu$bar at the poles, with the corresponding values of the eddy diffusion coefficient $K_{zz}$ ranging from $\sim$1000 m$^2$~s$^{-1}$ to 1-10 m$^2$~s$^{-1}$, respectively, at the 2~$\mu$bar level. This trend could be explained by upwelling at low latitudes and downwelling at high latitudes in both hemispheres and we estimate that vertical wind speeds of less than 2 cm~s$^{-1}$ would be required. The distributions of the photochemical products follow the homopause trend but they also show a clear seasonal trend at pressures between 0.01 and 10 $\mu$bar, with higher abundances in the summer hemisphere than in the winter hemisphere. We compare the observed distributions with results from one-dimensional seasonal photochemical models, with and without ion chemistry, to explore the impact of ion chemistry on the results. The best agreement between the models and the observations is obtained in the summer hemisphere, whereas disagreements in the winter hemisphere and auroral region may be due to the lack of transport by global circulation and auroral electron and ion precipitation in our photochemical models. Ion chemistry is particularly important for matching the observed C$_6$H$_6$ distribution, whereas differences between the neutral only and ion chemistry models are more modest for the other species. We also compare the C$_2$H$_2$ profiles retrieved from the UVIS occultations to those retrieved from the CIRS limb scans and find good agreement between the retrievals at pressures where they overlap. 

\end{abstract}


\end{frontmatter}


\section{Introduction}
\label{sc:intro}

\noindent
The middle and upper atmosphere plays an important role in shaping the composition and evolution of giant planets. The upper atmosphere separates the rest of the atmosphere from the extended magnetosphere and regulates the flow of material from external sources to the rest of the atmosphere \citep{Waite18,MOSES23}. Located above the tropopause and below the base of the thermosphere, from $\sim$~100 mbar to $\sim$~0.1--1 $\mu$bar, the middle atmosphere is the site of complex photochemistry that shapes the composition of the atmosphere \citep{MG05}. The 1970s saw advancements in our understanding of Saturn's middle atmosphere achieved through Earth-based observations and modeling \citep[e.g.,][]{Trafton73,Tokunaga75,Trafton77,CC79,MC79}. These studies supplied new information on the temperature, composition and seasonal behavior of Saturn's upper atmosphere, producing a basis for subsequent studies. In the 1980s, The Voyager missions provided higher spatial resolution observations, advancing our understanding of the composition, chemistry, dynamics, and energy balance of the middle and upper atmosphere of all of the outer planets \citep[e.g.,][]{Hanel81, Hanel83, Atreya84, Courtin84,Bishop90,Bishop92,Stevens93,Yelle93}. After its orbit insertion in 2004, the Cassini mission has provided a wealth of observations probing Saturn's middle and upper atmosphere, offering insights into seasonal and spatial trends in the atmosphere of this ringed giant planet \citep[e.g.,][]{Brown20,Koskinen21,Fletcher20}.

With its 26.7$^{\circ}$ axial tilt, Saturn experiences significant seasonality, magnified by the waning sunlight in the shadow of its extensive ring system in the winter hemisphere. Solar insolation is the first step in driving photochemistry in the middle and upper atmosphere, which also plays a role in powering dynamics in the middle atmosphere through its effect on thermal structure. Seasonal changes are also expected to be significant in the mesosphere, the region found between the stratosphere and the thermosphere where temperatures are relatively constant with height and which is characterized by relatively short photochemical and radiative timescales. The mesosphere is poorly understood and has not previously been probed by comprehensive observations that help to constrain chemistry and dynamics during a fixed season. The Cassini UVIS observations that we present in this work provide a unique view to this region.

Saturn's stratosphere and mesosphere are primarily composed of H\textsubscript{2}, He, H, and several light hydrocarbons, with CH$_4$ being the most abundant. Hydrocarbons other than methane are generated through photochemical processes, which are triggered by the solar ultraviolet radiation and in the auroral region, by energetic particle precipitation, causing the dissociation of CH$_4$. The hydrocarbons are expected to be confined to altitudes below the homopause where the atmosphere is well-mixed. Above the homopause, each species follows its own scale height and the abundances of species heavier than H$_2$ decrease rapidly with altitude. The pressure level of the homopause depends on mixing produced by atmospheric waves, turbulence \citep{Lindzen81}, and bulk transport properties of the atmosphere due to global circulation \citep[e.g.,][]{Fuller-Rowell98}. While photolysis of CH$_4$ serves as the primary source of minor hydrocarbons, the impact of this process on the overall abundance of CH$_4$ remains relatively modest and it does not drive the observed steep decrease in CH$_4$ abundance with altitude near the bottom of the thermosphere that is instead due to molecular diffusion. Observational constraints on the abundance of CH$_4$ as a function of latitude in the mesosphere can therefore act as a tracer of dynamics that constrains the mixing and circulation of the middle atmosphere.

Several previous studies have constrained the location of the homopause, although none of them provide coverage comparable to this work. The Voyager Ultraviolet Spectometer (UVS) instruments were used to observe three stellar and three solar occultations, a complete analysis of which was presented by \citet{VM15}. These authors defined the homopause as the pressure level where the volume mixing ratio of CH$_4$ decreases down to 5~$\times$~10$^{-5}$ (hereafter, the homopause reference level) from the stratospheric value of about 4.7~$\times$~10$^{-3}$ in order to avoid the uncertainties assotiated with the formal definition of the homopause as the location where the eddy and molecular diffusion coefficients are equal. In the three Voyager occultations probing the southern hemisphere at latitudes of 4.8--29$^{\circ}$ S, the homopause was located at 0.04-0.1 $\mu$bar while in the two occultations probing similar northern latitudes, the homopause was located closer to 0.01 $\mu$bar. The observations pointed to a possibility of a meridional trend or variation over time in the location of the homopause but did not provide sufficient coverage to confirm these trends. 

The Cassini UVIS instrument was used to observe dozens of stellar occultations during the 13-year in-orbit mission. Early work by \citet{SL12} used three of the occultations, at 43$^{\circ}$S, 4$^{\circ}$N and 15$^{\circ}$N planetocentric latitude, to place the homopause level at around 0.2 $\mu$bar, although they did not use the same definition as \citet{VM15}. The scope of this study was extended by \citet{KG18} who used Cassini Radio Science Subsystem (RSS) measurements of the lower atmosphere, temperature profiles retrieved from CIRS limb scans together with the density, temperature, and CH$_4$ profiles retrieved from UVIS occultations to construct atmospheric structure models for the occultation locations. The occultations, most of which were observed at planetocentric latitudes between 45$^{\circ}$S and 55$^{\circ}$N, showed considerable variability in the homopause reference level, which was centered at around 0.1 $\mu$bar. A single outlying point was observed with a deeper homopause near 1 $\mu$bar at 73$^{\circ}$ S planetocentric latitude. \citet{KG18} argued that this could be due to downwelling but were not able to draw clear conclusions due to the paucity of other observations at similarly high latitudes. Using Hubble Space Telescope (HST) images to probe the depth of auroral emission at Saturn, \citet{Gerard09} also found evidence for a deeper homopause at high latitudes than at low latitudes. The new observations presented in this work include several occulations at high latitudes and clearly confirm this trend. 

The scarcity of previous observations of the photochemical production region at high latitudes has also limited the investigation of the role of auroral processes on the photochemistry of the mesosphere and stratosphere. Studying the chemistry in this region is of interest since electron precipitation and ion chemistry can occur in the auroral regions and they are expected to play an important role in the formation of C$_6$H$_6$ and other higher order hydrocarbons \citep{Wong03, Vuitton08}. Forming in the mesosphere, C$_6$H$_6$ can undergo subsequent chemistry that produces polycyclic aromatic hydrocarbons (PAHs) and stratospheric haze, as suggested on Jupiter and Titan \citep{Friedson02,Wong02,Wong03,Vuitton08}. To address the formation of C$_6$H$_6$, \citet{Koskinen16} presented an analysis of the stellar occultations observed prior to 2016 and detected C$_6$H$_6$ at several latitudes, including at lower latitudes far from the auroral ovals. The authors compared their findings to a solar-driven neutral photochemical model and found that the observed C$_6$H$_6$ levels exceeded what the model could explain. They concluded that seasonally variable ion chemistry not included in the model was the primary driver of observed C$_6$H$_6$ levels, with a potential additional contribution from auroral ion chemistry at high latitudes and subsequent transport to lower latitudes. In order to demonstrate the feasibility of enhanced production by solar-driven ion chemistry, they used simple considerations based on a few reaction rate coefficients but did not pursue detailed modeling of the ion chemistry, as we do in this work.

\begin{figure}%
    \centering
    \noindent\includegraphics[width=\textwidth]{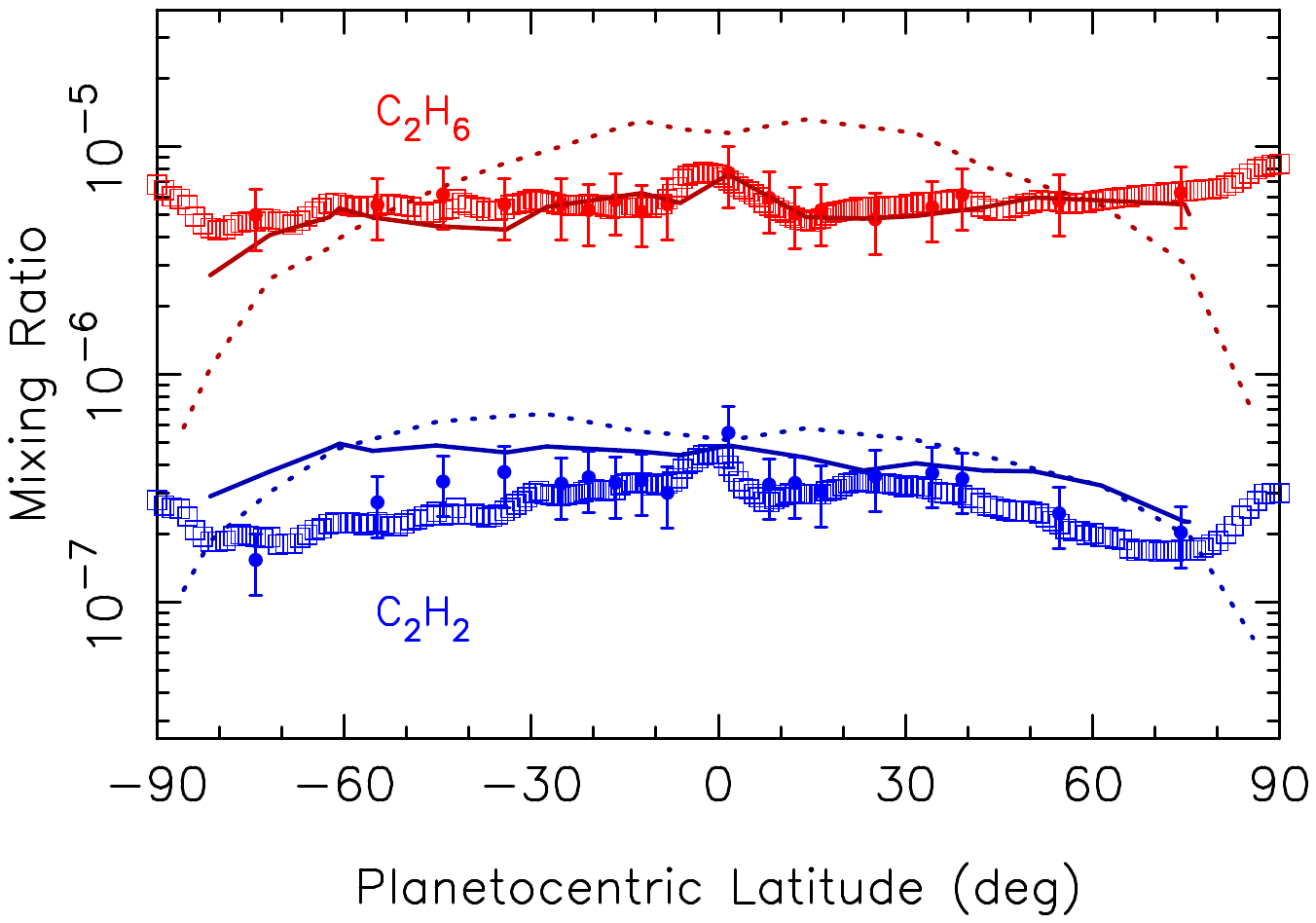}
    \caption{Volume mixing ratios of ethane (red) and acetylene (blue) at 1 mbar as a function of latitude, including Cassini Infrared Spectrometer (CIRS) limb scans from 2015 \citep[circles, this work and][]{Guerlet15} and nadir retrievals interpolated from a 2D grid of data observed on May 25, 2017 \citep[squares,][]{Fletcher20}. These observations are compared to results from two versions of a neutral photochemical model: an initial version where the K$_{zz}$ at each latitude was adjusted to reproduce the UVIS methane observations in the mesosphere and which used a constant K$_{zz}$ with latitude in the stratosphere (dotted lines), and a subsequent version featuring improved fits to CIRS stratospheric observations (thin solid lines). The latter version provides an improved approximation of the hydrocarbon levels at this pressure level, especially for ethane. These K$_{zz}$ profiles were used in the final iterations of the Neutral and Ion Models described in Section} \ref{M-JModel}.
    \label{fig:DiffKzz}
\end{figure}

Meridional trends in the abundances of the other hydrocarbon species can also constrain the chemistry and dynamics in the middle and upper atmosphere. Observations by Cassini/CIRS have allowed for the analysis of meridional trends at pressures greater than 0.01--0.1 mbar in the stratosphere \citep{fletcherbook18, Fletcher20}. The wide latitude coverage and frequent observation intervals relative to the Saturnian season also facilitate the monitoring of seasonal changes. The analysis of meridional trends has mostly focused on the distributions of C$_2$H$_6$, C$_2$H$_2$, methylacetylene (C$_3$H$_4$), diacetylene (C$_4$H$_2$), and propane (C$_3$H$_8$) that are readily observable in the CIRS data \citep{Howett07,Hesman09,Guerlet09,Guerlet10,FLETCHER09,FLETCHER15,Fletcher20,Sinclair13,Sinclair14}. Due to the relatively long lifetimes of C$_2$H$_6$ and C$_2$H$_2$, exceeding the Saturn year, any seasonal patterns in the middle and lower stratosphere (p $>$ 0.1 mbar) are damped \citep{MG05,Hue15,Hue16}. Under these conditions, photochemical models predict that the mixing ratios of both species should reach maximum around the equator and decrease with latitude towards the poles (see Figure \ref{fig:DiffKzz}). The CIRS observations broadly confirm both of these expectations for C$_2$H$_2$ \citep{Guerlet09, Sylvestre15,Fletcher20}, with the exception of an equatorial peak that could be due to downwelling associated with Saturn's quasi-periodic equatorial oscillation \citep{Guerlet09}. A similar peak is also seen in the distributions of C$_2$H$_6$ and C$_3$H$_8$, along with a peak in the retrieved temperatures that supports the interpretation in terms of localized downwelling around the equator \citep{Guerlet09,Sylvestre15}. It is possible that this could be related to Saturn's quasi-periodic oscillation, which alternates between upwelling and downwelling near the equator over time.

The meridional distributions of C$_2$H$_6$ and C$_3$H$_8$, on the other hand, vary less with latitude than expected. While there is no chemical explanation for the difference between longer-lived C$_2$H$_6$ and shorter-lived C$_2$H$_2$, meridional transport and associated vertical winds have been invoked to explain the distribution of C$_2$H$_6$ \citep{Greathouse05,Guerlet09,Sylvestre15}. In this context, it is interesting to note that the distribution of C$_2$H$_6$, in contrast to C$_2$H$_2$, does show substantial evolution with season at high latitudes around the polar vortices, decreasing during the fall and increasing during spring \citep{Fletcher20}. This could be consistent with seasonally changing dynamical drivers at middle to high latitudes. The situation for C$_3$H$_8$ is perhaps more confusing. \citet{Guerlet09} noted that its distribution appeared to align better with seasonal insolation than the distributions of the other two hydrocarbons. This, however, was not unambiguously demonstrated by the observed distribution, and the conclusion seems to have been contradicted by later observations pointing to limited seasonal evolution for the C$_3$H$_8$ distribution \citep{Sylvestre15}.

In the upper stratosphere (0.1--0.01 mbar), the photochemical timescales are shorter and the hydrocarbon distributions should begin to exhibit clear seasonal enhancement in the summer hemisphere and lower abundances in the winter hemisphere. Curiously, the C$_2$H$_2$ and C$_2$H$_6$ abundances retrieved from CIRS limb scans at the the 0.01 mbar level do not agree with this expectation at all. Instead, both distributions included a deep minimum at latitudes of 25--45$^{\circ}$S and a maximum at around 25$^{\circ}$N \citep{Guerlet09} in the winter hemisphere. These features could be explained by upwelling circulation in the southern (summer) hemisphere and downwelling circulation in the northern (winter) hemisphere, which would also explain the lack of a clear signature of the ring shadow in the temperature observations in the northern hemisphere \citep{Guerlet09, Guerlet14}. In subsequent limb scans observed in 2010 and 2012 during early northern spring, \citet{Sylvestre15} noted that the local minimum at southern low latitudes had remained relatively stable. However, the local maximum near 25$^{\circ}$N was not apparent in these later observations. These authors highlighted that, in contrast to the temperature patterns in the spring, winter temperatures were observed to be warmer, and summer temperatures were cooler than model predictions \citep{FM12}. This discrepancy led the authors to suggest that the influence of dynamics might be more pronounced during winter than in spring. The evidence from CIRS spanning from late northern winter to early northern spring points to the possibility of seasonally reversing meridional circulation in the upper stratosphere. The seasonal evolution of such a circulation cell would need to include a time lag on the order of one Saturn season to remain consistent with the observations\citep{Moses05,Sylvestre15}.

Several models and other studies are also consistent with seasonally changing meridional circulation in the stratosphere. For example, \citet{FM12} used the three-dimensional Outer-Planet General Circulation Model (OPGCM) to predict that stratospheric circulation at low latitudes is dominated by seasonally reversing Hadley circulation, characterized by meridional flows from the summer to the winter hemisphere, with strong subsidence around 25$^{\circ}$ latitude in the winter hemisphere. In more recent work, \citet{Bardet22} used the DYNAMICO-Saturn GCM to characterize the evolution of the equatorial oscillation and interhemispheric circulation in Saturn's stratosphere. They predict an inter-hemispheric (summer to winter) cell under solstice conditions and a single, common, upwelling branch centered at the equator and two downwelling branches located around 40$^{\circ}$ latitude in either hemisphere at equinox conditions. These meridional flows are consistent with seasonal forcing, which are observed to transition from one hemisphere to the other around the equinox \citep{Fletcher20}. Taken together, previous observations and models of the stratosphere therefore lead to the expectation of upwelling near the sub-solar latitude and meridional flows towards the winter hemisphere and possibly towards the summer pole at the time of the northern summer solstice.

While meridional flows in the stratosphere can be maintained by seasonally changing solar insolation, at higher altitudes in the thermosphere solar heating is negligible and electric currents associated with the aurora appear to provide the main energy and momentum sources, along with gravity waves \citep{Mueller-Wodarg06,Mueller-Wodarg12,Mueller-Wodarg19,Sylvestre15,Baines18,Brown20,Brown22}. \citet{Brown20} analyzed Cassini UVIS stellar occultations obtained during the Grand Finale tour to map temperatures and pressure levels in the thermosphere and used the results to infer circulation at around the northern summer solstice. The temperatures peak at low pressures near the auroral latitudes. Based on the observed atmospheric structure, they inferred divergent meridional winds at these latitudes, which are needed to help explain the higher-than-expected temperatures throughout the thermosphere. The proposed circulation cell would consist of upwelling near auroral latitudes and subsequent flow towards the equator and poleward from the auroral region, in both hemispheres. This circulation is opposite to the meridional flows inferred for the stratosphere below. At the same time, there is a steep gradient in temperature between the poles and the equator at pressures greater than the 0.01 $\mu$bar level in the lower thermosphere, with an effectively deeper base of the thermosphere at the poles than at the equator. This could be consistent with meridional flows away from the subsolar region in the mesosphere and lower thermosphere, as long as equatorward circulation exists at lower pressures. In line with this idea, \citet{VM15} also suggest the presence of a circulation cell in the mesosphere with low-latitude upwelling that would be consistent with a lower-pressure homopause near the subsolar latitude. There are, however, almost no observational constraints on the circulation in the mesosphere in previous work that would allow us to firmly connect the dynamics in the stratosphere to that of the lower thermosphere. 

In this work, we use observations from the last two years of the Cassini mission to give the first comprehensive view of Saturn's middle and upper atmosphere around the northern summer solstice. We analyze and interpret stellar occultations observed by the UVIS instrument in 2016 and 2017, including the Cassini Grand Finale occultations obtained during a 6-week time period in the late summer of 2017, shortly before the end of mission. The 33 occultations span latitudes from pole to pole and include several that probe high latitudes near the auroral ovals (see Table~\ref{table:UVISdata}). The density and temperature in the thermosphere from these occultations was already retrieved by \citet{Brown20}. In this work, we present densities of CH$_4$, C$_2$H$_6$, C$_2$H$_2$, C$_2$H$_4$, and C$_6$H$_6$ in the upper stratosphere and mesosphere retrieved from the occultations. As opposed to the deeper pressure levels in the stratosphere, seasonal trends in the hydrocarbon abundances are expected to be detectable in this region. We also detect C$_6$H$_6$ in most occultations and by comparing the observed C$_6$H$_6$ abundances to model predictions, help to constrain the importance of ion chemistry in producing this molecule in Saturn's atmosphere. Because the observations span the auroral region, we can probe the possible impact of auroral processes on the chemistry. In addition, meridional trends that depart from expectations based on 1D photochemical models provide some of the only constraints on dynamics in the upper stratosphere and mesosphere.

We combine the UVIS observations with CIRS limb scans of the stratosphere to provide a full picture of the middle and upper atmosphere with pressure and latitude. The CIRS limb scans used in this work were obtained between 2015 and 2017 (see Table~\ref{table:CIRSdata}). Some of them were used in previous work \citep{Guerlet18,KG18,MOSES23} while others have not been published before. The limb scans can be used to retrieve the pressure, temperature, and abundances of several light hydrocarbons, including those of C$_2$H$_2$ and C$_2$H$_6$, which are also retrievable from the UVIS occultations. Combining the results of both analyses, we create a temperature map of the middle and upper atmosphere and provide constraints on the C$_2$H$_2$ and C$_2$H$_6$ abundances as a function of latitude and depth, spanning from the lower stratosphere to above the homopause level. We also produce atmospheric structure models that provide pressure, temperature, total density, and altitude for the occultation locations. These results support physical models of the atmosphere and future mission planning. We then compare the retrieved hydrocarbon distributions to results from a seasonal photochemical model. To emphasize ion chemistry's influence on hydrocarbon distributions, we compare models with and without ion chemistry and explicitly include it in the photochemical model, offering a comprehensive view of its impact on observable hydrocarbons. 

This paper is organized as follows. We begin with a description of the methods used to retrieve the hydrocarbon distributions from the UVIS data in Section~\ref{M-UVIS} and the temperatures and hydrocarbon distributions from the CIRS data in Section~\ref{M-CIRS}. In order to compare the UVIS results to our photochemical model, which we describe in Section~\ref{M-JModel}, we derive an atmospheric structure model, which we describe in Section~\ref{M-Atmo}. We present the key results of this study in Section~\ref{Results}. These include a meridional trend in the location of the homopause in Section~\ref{R-Homo}, a comparison of the observed hydrocarbon distributions with the photochemical model in Section~\ref{R-Hydro} as well as a separate discussion of the distribution of C$_6$H$_6$ and ion chemistry in Section~\ref{R-Benzene}. We then discuss these implications of these results to circulation and dynamics in Section~\ref{D-Homo}, the seasonal behavior of hydrocarbons in Section~\ref{D-Hydro}, and the implications for auroral chemistry in Section~\ref{D-Auroral}.

\section{Methods}

\noindent
As stated above, this work uses temperatures in the stratosphere and the  thermosphere together with hydrocarbon densities in the stratosphere and mesosphere retrieved from stellar occultations observed by the Cassini UVIS instrument and limb scans observed by the Cassini CIRS instrument. Below, we summarize the retrieval methods, describe the methods to create the atmospheric structure models, and give details of the photochemical model.

\subsection{UVIS FUV Hydrocarbon Retrievals} \label{M-UVIS}

\noindent
The FUV channel probes wavelengths between 1115 and 1912 \AA~where the light hydrocarbons CH$_4$, C$_2$H$_6$, C$_2$H$_4$, C$_2$H$_2$ and C$_6$H$_6$ have absorption features. These observations probe pressures between $10^{-5}$ and 0.1 mbar, depending on the species and occultation. We summarize the data processing and retrieval of the hydrocarbons from the FUV channel data below; further details can be found in the supplementary information to \citet{Koskinen16}. Details of the H\textsubscript{2} density and temperature retrieval from the EUV channel data for these occultations are described by \citet{Brown20}.

\begin{table}
\centering
\caption{UVIS FUV Stellar Occultation}
\begin{tabular}{ ccccc } 
\hline
 Occultation ID & Date & Target & Lat[deg] & LST [hrs] \\
\hline

FUV2016\textunderscore014\textunderscore21\textunderscore05 & 2016-Jan-14 & $\epsilon$ Ori & 25.4 & 19.2 \\

FUV2016\textunderscore045\textunderscore02\textunderscore15 & 2016-Feb-13 & $\alpha$ Vir & 4.4 & 2.6  \\

FUV2016\textunderscore045\textunderscore08\textunderscore50 & 2016-Feb-14 & $\alpha$ Vir &  1.5 & 14.8 \\

FUV2016\textunderscore046\textunderscore23\textunderscore30 & 2016-Feb-15 & $\gamma$ Ori & 66.1 & 4.6 \\

\textsuperscript{\textdagger}FUV2016\textunderscore094\textunderscore18\textunderscore01 & 2016-Apr-03 & $\gamma$ Ori & 0.6 & 18.7 \\

FUV2016\textunderscore296\textunderscore06\textunderscore40 & 2016-Oct-22 & $\zeta$ Cru & 14.4 & 16.3 \\

FUV2016\textunderscore339\textunderscore04\textunderscore47 & 2016-Dec-03 & $\beta$ Cru & -6.2 &  2.2  \\

FUV2016\textunderscore353\textunderscore13\textunderscore13 & 2016-Dec-18 & $\beta$ Cru & -6.0 &  2.1  \\

FUV2017\textunderscore009\textunderscore00\textunderscore55 & 2017-Jan-08 & $\alpha$ Cru & -13.1 &  3.5 \\

FUV2017\textunderscore087\textunderscore21\textunderscore53 & 2017-Mar-28 & $\beta$ Cru & -5.0 &  1.7  \\

\textsuperscript{\textdagger}FUV2017\textunderscore139\textunderscore19\textunderscore57\textunderscore58 & 2017-May-19 & $\alpha$ CMa & -65.4 &  5.7  \\

\textsuperscript{\textdagger}FUV2017\textunderscore146\textunderscore05\textunderscore39\textunderscore00 & 2017-May-25 & $\alpha$ CMa & -73.3 &  5.7  \\

\textsuperscript{$\ast$}FUV2017\textunderscore175\textunderscore20\textunderscore36\textunderscore24 & 2017-May-19 & $\epsilon$ Ori & -82.4 & 7.5 \\

\textsuperscript{$\ast$}FUV2017\textunderscore176\textunderscore06\textunderscore40\textunderscore50 & 2017-Jun-25 & $\epsilon$ Ori & 12.2 &  5.8  \\

\textsuperscript{\textdagger}FUV2017\textunderscore176\textunderscore11\textunderscore17\textunderscore50 & 2017-Jun-25 & $\zeta$ Ori & 23.9 &  5.7  \\

\textsuperscript{$\ast$}FUV2017\textunderscore182\textunderscore06\textunderscore15\textunderscore00 & 2017-Jul-01 & $\epsilon$ Ori & -85.7 & 12.4  \\

\textsuperscript{\textdagger}FUV2017\textunderscore182\textunderscore08\textunderscore43\textunderscore29 & 2017-Jul-01 & $\zeta$ Ori & -78.8 &  0.8  \\

\textsuperscript{$\ast$}FUV2017\textunderscore182\textunderscore20\textunderscore40\textunderscore30 & 2017-Jul-01 & $\epsilon$ Ori & 31.7 &  5.7 \\

\textsuperscript{$\ast$}FUV2017\textunderscore183\textunderscore00\textunderscore47\textunderscore29 & 2017-Jul-01 & $\zeta$ Ori & 42.5 & 5.7 \\

\textsuperscript{$\ast$}FUV2017\textunderscore188\textunderscore17\textunderscore43\textunderscore18 & 2017-Jul-07 & $\epsilon$ Ori & -71.9 & 17.0 \\

\textsuperscript{$\ast$}FUV2017\textunderscore188\textunderscore20\textunderscore24\textunderscore30 & 2017-Jul-07 & $\zeta$ Ori & -62.3 & 17.5 \\

\textsuperscript{\textdagger}FUV2017\textunderscore189\textunderscore09\textunderscore54\textunderscore30 & 2017-Jul-08 & $\epsilon$ Ori & 49.6 & 5.5 \\

\textsuperscript{$\ast$}FUV2017\textunderscore194\textunderscore14\textunderscore52\textunderscore10 & 2017-Jul-13 & $\gamma$ Ori & -34.1 & 17.0 \\

\textsuperscript{$\ast$}FUV2017\textunderscore195\textunderscore02\textunderscore00\textunderscore00 & 2017-Jul-13 & $\gamma$ Ori & 73.5 & 17.4 \\

\textsuperscript{$\ast$}FUV2017\textunderscore195\textunderscore05\textunderscore17\textunderscore19 & 2017-Jul-13 & $\epsilon$ Ori & -55.4 & 17.4 \\

\textsuperscript{$\ast$}FUV2017\textunderscore195\textunderscore08\textunderscore53\textunderscore10 & 2017-Jul-14 & $\zeta$ Ori & -45.2 & 17.7 \\

\textsuperscript{\textdagger}FUV2017\textunderscore196\textunderscore00\textunderscore53\textunderscore59 & 2017-Jul-14 & $\zeta$ Ori & 81.4 & 5.2 \\

\textsuperscript{$\ast$}FUV2017\textunderscore201\textunderscore21\textunderscore41\textunderscore00 & 2017-Jul-20 & $\zeta$ Ori & -27.5 & 17.8 \\

\textsuperscript{$\ast$}FUV2017\textunderscore202\textunderscore09\textunderscore04\textunderscore00 & 2017-Jul-217 & $\epsilon$ Ori & 85.7 & 0.3 \\

\textsuperscript{$\ast$}FUV2017\textunderscore202\textunderscore11\textunderscore31\textunderscore59 & 2017-Jul-21 & $\zeta$ Ori & 74.7 & 18.7 \\

FUV2017\textunderscore202\textunderscore21\textunderscore21\textunderscore50 & 2017-Jul-21 & $\kappa$ Ori & -80.0 & 20.5 \\

\textsuperscript{$\ast$}\textsuperscript{\textdagger}FUV2017\textunderscore209\textunderscore09\textunderscore39\textunderscore55 & 2017-Jul-28 & $\kappa$ Ori & -60.7 & 18.7 \\

\textsuperscript{$\ast$}FUV2017\textunderscore210\textunderscore04\textunderscore22\textunderscore00 & 2017-Jul-29 & $\kappa$ Ori & 61.4 & 6.7 \\

\hline
\end{tabular}
\begin{tablenotes}
\scriptsize
\item Local solar time and planetocentric latitude and longitude are taken at the half light point. Latitudes are planetocentric.
\item\textsuperscript{$\ast$}~These observations were used to create latitude-specific 1D models of hydrocarbon abundances using the Ion and Neutral photochemical models.
\item\textdagger~These occultations lacked a corresponding EUV retrieval. The atmospheric structure cannot be calculated without these thermospheric temperature profiles. Therefore number densities cannot be converted to mixing ratios and they are not included in comparisons to the photochemical models.
\end{tablenotes}
\label{table:UVISdata}
\end{table}

During each stellar occultation, the UVIS detector recorded the spectrum of a UV-bright star as the line of sight from the star to the detector passed through Saturn's atmosphere. The occultations were observed simultaneously in the EUV and FUV channels, probing the hydrocarbons in the mesosphere and the H\textsubscript{2} density and temperature in the thermosphere. The exposure times varied from 1.75 to 3.125 s in the FUV and from 3.00 to 8.75 s in the EUV, producing altitude resolutions of 2 to 29 km in the FUV and 13 to 53 km in the EUV. The spatial span of a single occultation varied from 0.004$^{\circ}$ to 9.95$^{\circ}$ in latitude and from 2.14$^{\circ}$ to 31.04$^{\circ}$ in longitude over the entire altitude span covered by both detectors (about 2,500 km). In most cases, the latitude span of individual occultations was less than 1$^{\circ}$ and the longitude span less than 10$^{\circ}$ and the path length at the 1 $\mu$bar pressure level near the center of the occultations is less than 5,500 km, which justifies the treatment of the occultations as vertical scans of the atmosphere at a given latitude and longitude. In the few cases where this may be questionable, we still assume that horizontal variations are sufficiently small to justify this approach. The occultations were observed at a range of longitudes and local solar times, with many of them near dawn and dusk. 

We retrieved the line of sight (LOS) column densities of the hydrocarbons as a function of tangent altitude by using the Levenberg-Marquardt (L-M) algorithm \citep{Markwardt09} to fit a forward model to the observed transmission spectrum separately for each LOS through the atmosphere. The transmission spectra were calculated by dividing the stellar signal transmitted through the atmosphere by the unocculted stellar spectrum, which was obtained by averaging the unocculted spectra over a range of high altitudes where absorption by the atmosphere is negligible. Tangent altitudes are defined as the distance along the surface normal to the LOS at the closest approach from the 1 bar level. We used the shape model from the Saturn Atmospheric Modeling Working Group for the 1 bar and 0.1 bar levels \citep{SAMWGReport}. For the gravitational potential in this work, we used the Cassini fit from \citet{AS07}. This potential incorporates the J$_{2n}$ numerical coefficients, the P$_{2n}$ Legendre polynomials, and the rotational potential to account for perturbations to spherical symmetry in Saturn's gravity field. As shown by \citet{Koskinen21}, this gravity field is sufficiently similar to the gravity field with differential rotation based on the Cassini Grand Finale measurements \citep{Militzer19} to be suitable for our purposes. 

Prior to retrieving the LOS column densities, we processed the data and checked for evidence of pointing offsets and/or background contamination. For these observations, the image of the star was centered in one or two spatial pixels of the detector, with residual signal at adjacent pixels. We summed the signal over four spatial pixels to obtain the transmitted spectra as a function of tangent altitude and wavelength. Then we investigated the behavior of the light curves (transmission vs. tangent altitude) in 50 \AA~bins and looked for evidence of pointing drifts and other artifacts. In agreement with the pointing information in the telemetry, the light curves are consistent with stable observations without significant pointing drifts. However, we excluded six occultations that showed evidence of ramp-like behavior at altitudes between 700 and 3,000 km, likely due to detector adjustment to a bright star during egress occultations \citep{Koskinen16}.

We detected a small amount of background light arising from Saturn shine and scattered light in the instrument.The background ranged from 0.05\% to 18\% of the unocculted stellar signal at 1250--1900 \AA~in 50 \AA~wavelength bins per integration period, with a median value of 0.4\%. To correct for this, we checked each of the thirteen 50 \AA~wavelength bins between 1250 and 1900~\AA~at altitudes where the signal was flat and the star was fully occulted. Once the altitudes with no stellar signal were identified (between 100 and 320 km above the 1 bar level, with an average extent of 196 km), the average signal in each individual 0.78 \AA~wavelength band at these altitudes was subtracted from the signal at all altitudes. We then designated altitude regions for the unocculted stellar reference spectrum where we averaged signal in each wavelength band to obtain the reference spectra for transmission. 

Finally, we designated ranges over which to retrieve column densities by finding the altitudes where the observed optical depth was between 0.1 and 10 over a wavelength range of at least 300 \AA. In addition, the wavelength calibration depends on the position of the star in the field of view. In order to correct for any offsets from the true wavelengths of the observed absorption, we introduced small shifts to the wavelength calibration for each occultation, on the order of 0.5 \AA, as appropriate. This constant offset for each occultation was determined by cross-correlating the model transmission spectrum with the observed transmission spectrum at altitudes roughly commensurate with the retrieval altitudes and generally between 300 and 930 km. Because the pointing was stable during these occultations, we used the same shift at every altitude. The wavelength calibration is aided by strong C$_2$H$_2$ and C$_2$H$_4$ absorption bands in the transmission spectrum, which are clearly evident in all occultations and are described later in this section.

Observed transmission depends on the optical depth due to gaseous absorbers and the response function of the detector. We also looked for evidence of aerosol absorption in these occultations but did not find any. \citet{Koskinen16} identified evidence for aerosol extinction for one occultation at 73$^\circ$S latitude and concluded that this was likely due to the condensation of benzene in a particularly cold location during polar night. However, our observations in this same shadowed region did not show evidence of such extinction.

Having retrieved the LOS column densities of the absorbers, we multiplied the uncertainties on the column densities based on the L-M algorithm by a factor of three, guided by the simulated occultation retrieval of \citet{Koskinen16}. We then inverted the retrieved column densities using Tikhonov regularization to obtain local number densities as a function of altitude \citep{Koskinen11}. Tikhonov regularization is a numerical technique suitable for ill-posed inversion problems, stabilizing solutions against small changes in initial data by reducing uncertainty and effectively smoothing densities with altitude \citep{TA77}. The final altitude resolutions of our retrieved density profiles depend on the Tikhonov regularization parameter used. While this resolution is lower than the initial sampling resolution, the altitude resolution of the results is still relatively good. We used Tikhonov parameters between 0.01 and 2, depending on the occultation and species, to achieve tolerable error bars while preserving an acceptable vertical resolution. For example, the altitude resolution for the CH$_4$ retrieval ranges from 6 to 46 km, with an average vertical resolution of 14 km. The altitude resolutions and average values for the other species are 10 km to 22 km, 13 km and C$_2$H$_2$; 10 km to 33 km, 17 km and for C$_2$H$_4$; 9 km to 20 km, 13 km and for C$_2$H$_6$; and 10 km to 36 km and for C$_6$H$_6$.

\begin{figure}
\noindent\includegraphics[width=\textwidth]{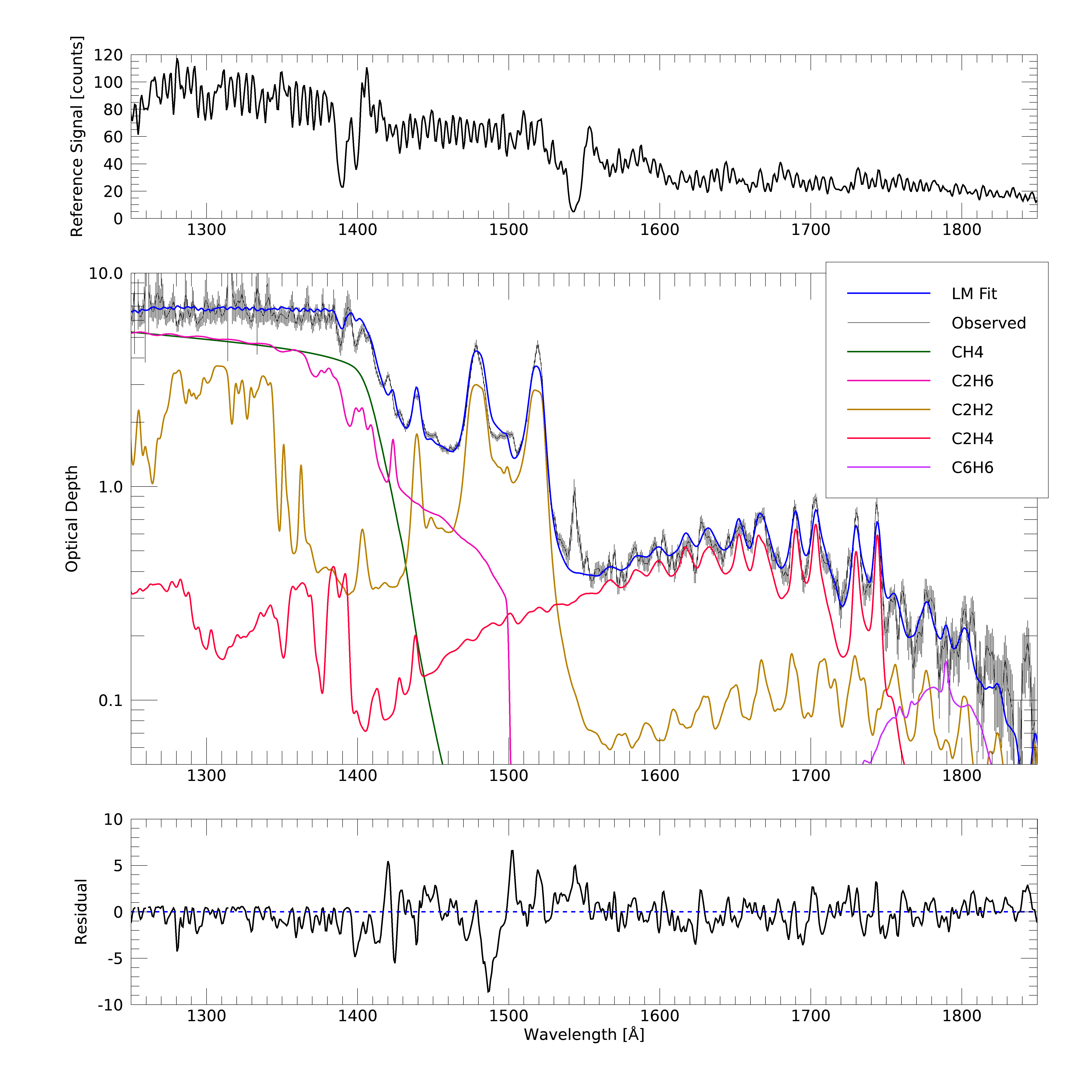}
\caption{Transmission spectrum for an occultation of Epsilon Orionis (FUV2017\textunderscore182\textunderscore20\textunderscore40\textunderscore30 in Table~\ref{table:UVISdata}) probing 32$^{\circ}$ N planetocentric latitude is shown in the center panel. The observed transmission was binned between 675 and 725 km altitude and converted to optical depth. The observed optical depth is shown in grey with error bars and the forward model fit is shown in blue (center). The contributions to optical depth from each species are shown separately (CH$_4$ green, C$_2$H$_2$ gold, C$_2$H$_4$ red, C$_2$H$_6$ pink, C$_6$H$_6$ purple) based on the best fit column density and molecular cross sections. The top panel shows the stellar reference spectrum. The bottom panel shows the residuals to the fit.}
\label{fig:transmission}
\end{figure}

Prior to the retrieval, we checked every occultation for absorption by each of the five species and included a species in the final retrieval only if both a) it improved the fit to transmission at one or more binned 50 km intervals and b) the fit residual decreased at wavelengths where the species' absorption cross section is large. Since the temperature variations in the upper stratosphere and mesosphere are not large enough to necessitate the use of temperature-dependent cross sections, our analysis used constant, low-temperature cross sections for CH$_4$ (150 K) \citep{ChenWu04}, C$_2$H$_2$ (150 K) \citep{Wu01}, C$_2$H$_4$ (140 K) \citep{Wu04}, C$_2$H$_6$ (150 K) \citep{ChenWu04}, and C$_6$H$_6$ (215 K) \citep{Capalbo16}. Owing to the relative abundances of the hydrocarbons and the characteristics of the cross sections, the vertical coverage of each occultation is species-dependent. An example of the L-M fit to the observed optical depth is shown in Figure \ref{fig:transmission}. Here, we binned the observed transmission at tangent altitudes of 570--620 km for the occultation of Kappa Orionis at the planetocentric latitude of 61$^{\circ}$ N in order to clearly show each of the species present in the transmission spectrum.

The CH$_4$ cross section is relatively flat in the 1100--1400\AA~ wavelength region where it dominates absorption; therefore CH$_4$ densities are only retrieved over a small pressure range. Lower boundaries of the retrieval for CH$_4$ vary between 0.4 and 15 $\mu$bar at different latitudes. Similarly, the C$_2$H$_6$ cross section is only significant at wavelengths shorter than $\sim$1500 \AA, but the cross section from approximately 1350 to 1500 \AA~allows the C$_2$H$_6$ retrieval to extend somewhat deeper than the CH$_4$ retrieval. Separating the contributions of CH$_4$ and C$_2$H$_6$ is tricky in stellar occultations by Titan's atmosphere \citep{Koskinen11} but on Saturn, the separation is clearer, given the lack of aerosol extinction and additional hydrocarbon and nitrile absorbers found in Titan's atmosphere \citep{Koskinen16}.

The strong absorption bands of C$_2$H$_2$ near 1500 \AA~are easily fit at equatorial altitudes between 700 and 900 km. Absorption by these two bands saturates at lower altitudes but other spectral features at 1600 to 1900 \AA~are still detectable. Thus,  C$_2$H$_2$ is observable over a relatively large pressure range (approximately 200 to 0.01 $\mu$bar, see Figure \ref{fig:ndensprofile}). Ethylene was observed over a similar pressure range (100 to 0.04 $\mu$bar), due to its strong absorption bands longward of 1550 \AA. Benzene has a large cross section near 1800 \AA, which allows it to be distinguished from C$_2$H$_4$ and C$_2$H$_2$, even though those molecules are present in much larger quantities \citep{Koskinen16}. Benzene is detected in 29 of 33 occultations; between 15$^{\circ}$S and 86$^{\circ}$N planetocentric latitudes, it is detected between 1 $\times$ 10\textsuperscript{-8} and 1 $\times$ 10\textsuperscript{-5} bar. Benzene is not detected at 12$^{\circ}$N planetocentric latitude or in three occultations probing the atmosphere between planetocentric latitudes of 27$^{\circ}$S and 45$^{\circ}$S. In the southern (winter) hemisphere, we detect C$_6$H$_6$ between 55$^{\circ}$S and 86$^{\circ}$S planetocentric latitudes over a smaller range of pressures than in the north, between 600 to 2 $\mu$bar, due to the deeper homopause pressure in the south (see Section \ref{R-Homo}).

In some transmission spectra, we observed an unusual feature near 1544~\AA \\(see Figure \ref{fig:transmission}) that was not well-matched by our model. We considered the possibility that a previously unnoticed species could be contributing to the absorption at this wavelength. The most plausible absorber that we found is carbon monoxide (CO); the CO fourth positive system has a strong absorption feature around this wavelength. Including CO in our simultaneous fits, however, led to the prediction of a second strong line, the (1-0) band of the CO fourth positive system, near 1510 \AA, which is not observed in the data. It is possible that the feature near 1544 \AA~in the occultation data is an artifact caused by a strong absorption feature in the stellar spectrum. There is a deep and broad absorption feature centered near 1544 \AA~in the spectrum of the supergiant stars Epsilon Orionis, Kappa Orionis and Zeta Orionis, used for occultations in which this feature is strongest. The stellar flux in the spectrum of Epsilon Orionis, in particular, drops to near zero around 1544 \AA~(see Figure \ref{fig:transmission}). In this case, the wings of the line spread function of the instrument distribute the signal from other wavelengths to this feature. The retrieval of column density from transmission would ideally be performed with a stellar signal that is constant with wavelength. When the stellar signal is very close to zero, the calculation of transmission is confounded and may result in artifacts that mimic absorption features, as observed here.

\subsection{CIRS Limb Scans}
\label{M-CIRS}

\noindent
A total of 25 CIRS limb scans were included in this work, covering a period from June 2015 to August 2017 and spanning planetographic latitudes from 77$^{\circ}$ S to 77$^{\circ}$ N (see Table 2). Among these scans, temperature profiles retrieved from the 2015 observations were previously published by \citet{Guerlet18} and \citet{KG18}. The remaining limb scans, obtained between November 19, 2016, and August 26, 2017, are analyzed and presented in this study for the first time along with previously unreleased hydrocarbon profiles from all the 2015 observations. The methods used to retrieve the temperature profiles are described in detail by \citet{Guerlet09} and \citet{Sylvestre15}, and only a brief summary is provided here. CIRS is a Fourier transform spectrometer that has three focal planes, of which we used two: the FP3 and FP4 focal planes. The FP3 focal plane covers the wavelengths of 9 to 17 $\mu$m and probes the H$_2$-H$_2$ and H$_2$-He collision-induced continuum emissions (probing temperatures at pressures from about 1 to 10 mbar) as well as the $\nu_{9}$ emission band of C$_2$H$_6$ and the $\nu_{5}$ emission band of C$_2$H$_2$. The FP4 focal plane covers wavelengths of 7 to 9 $\mu$m, including CH$_4$ emissions that probe temperatures at pressures lower than $\sim$5 mbar and up to a few $\mu$bar. During limb scans, the parallel arrays of FP3 and FP4 were set approximately perpendicular to the limb of the planet. The ten square detectors were arranged linearly in a rectangular array, with individual projected fields of view of approximately 60 kilometers, which correspond to about 1 to 1.5 atmospheric scale heights. 

\begin{table}
    \centering
        \begin{threeparttable}
                \caption{CIRS Limb Scan Data}
                    \begin{tabular}{ cc } 
                    \hline
                    Date & Planetographic Latitude [$^{\circ}$] \\
                    \hline
                    \textsuperscript{\textdagger}6/16/2015 & 77S, 40S, 25S, 15S, \textsuperscript{\textdagger}2N, 15N, 40N, 77N \\
                    \textsuperscript{\textdagger}9/30/2015 & 60S, 50S, 30S, 20S, 10S \\
                    \textsuperscript{\textdagger}11/24/2015 & 10N, 20N, 30N, 45N, 60N \\
                    11/19/2016 & 15S, 0 \\
                    01/16/2017 & 5S \\
                    03/07/2017 & 10S, 10N \\
                    08/13/2017 & 5N \\
                    08/26/2017 & 0\\
                    \hline
                    \end{tabular}
                    \begin{tablenotes}
                    \small
                    \item\textdagger~Temperature profiles observed on these dates were previously published in \citet{Guerlet18} and \citet{KG18}. All hydrocarbon retrievals from these observations presented here are new results. 
                    \end{tablenotes}
                    \label{table:CIRSdata}
    \end{threeparttable}
\end{table}

A radiative transfer model and a Bayesian retrieval algorithm are used to retrieve the vertical temperature profile from the scans. The algorithm also applies a correction to the a priori LOS tangent altitudes to obtain a better fit to the data \citep{Guerlet09}. The altitude correction is necessary because the a priori altitudes are affected by the CIRS pointing uncertainty and are based on the NAIF reference ellipsoid that differs substantially from the true 1 bar level shape \citep{KG18}. The limb scans record simultaneous spectra at different tangent altitudes and include the saturation (peak) altitude of the emission profile that depends strongly on the pressure along the LOS. Therefore, they constrain the altitudes of the pressure levels in the atmosphere, in addition to the temperature.

The analysis begins with the retrieval of a vertical temperature profile from the $\nu_{4}$ band of CH$_4$ (from 1200 to 1370 cm$^{-1}$) that probes the middle and upper stratosphere and from the H\textsubscript{2}-H\textsubscript{2} and H\textsubscript{2}-He collision-induced emission (from 590 to 660 cm$^{-1}$) that probe the lower stratosphere. The analysis uses regularization to smooth the profiles to filter out spurious small-scale oscillating patterns (smaller than one scale height, unresolved by the limb data) and an a priori constraint \cite[see][for a description of the retrieval algorithm]{Guerlet09}. For a given data set, the best-fit solution is chosen among several retrievals that start from two different a priori temperature profiles and typically four sets of regularization parameters. At pressures greater than 20 mbar, the limb temperature retrieval converges to the a priori profile derived from Voyager radio occultations \citep{Lindal85}, setting the high-pressure extent of the observation.

The upper boundary of temperature retrieval using CH$_4$ emissions is determined by the signal-to-noise of the spectra and varies across observations. The observed emissions may also be affected by non-LTE conditions. In previous CIRS analyses \citep[e.g.,][]{Guerlet09,Guerlet11,Guerlet18}, the Planck function provided a good approximation to the source function up to the 1 $\mu$bar level for the 7.7 $\mu$m CH$_4$ band, as was suggested by \citet{Appleby90}. However, in five of the dayside limb scans presented here, we observed spectral features in the CH$_4$ band that could not be fitted by our LTE radiative transfer model at tangent altitudes corresponding to pressures of 10 or even 100 $\mu$bar, indicating possible non-LTE emissions. We therefore limited these observations to pressures greater than 0.3 mbar, despite good SNR ratio. As a result, the lower pressure boundaries for the CIRS temperature retrieval across all latitudes varies from about 0.3 to 0.005 mbar. The high-pressure limit of the UVIS temperature profiles, on the other hand, extends down to the 0.01 $\mu$bar level near the equator and the 1 $\mu$bar level near the poles. Hydrocarbon volume mixing ratios (VMRs) are retrieved using a similar retrieval algorithm by setting the temperature to that previously retrieved and by fitting the $\nu_{9}$ emission band of C$_2$H$_6$ centered at 822 cm$^{-1}$ and the $\nu_{5}$ emission band of C$_2$H$_2$ centered at 730 cm$^{-1}$ in FP3.

\subsection{Atmospheric Structure Model}
\label{M-Atmo}

\begin{figure}
\noindent\includegraphics[width=0.75\textwidth]{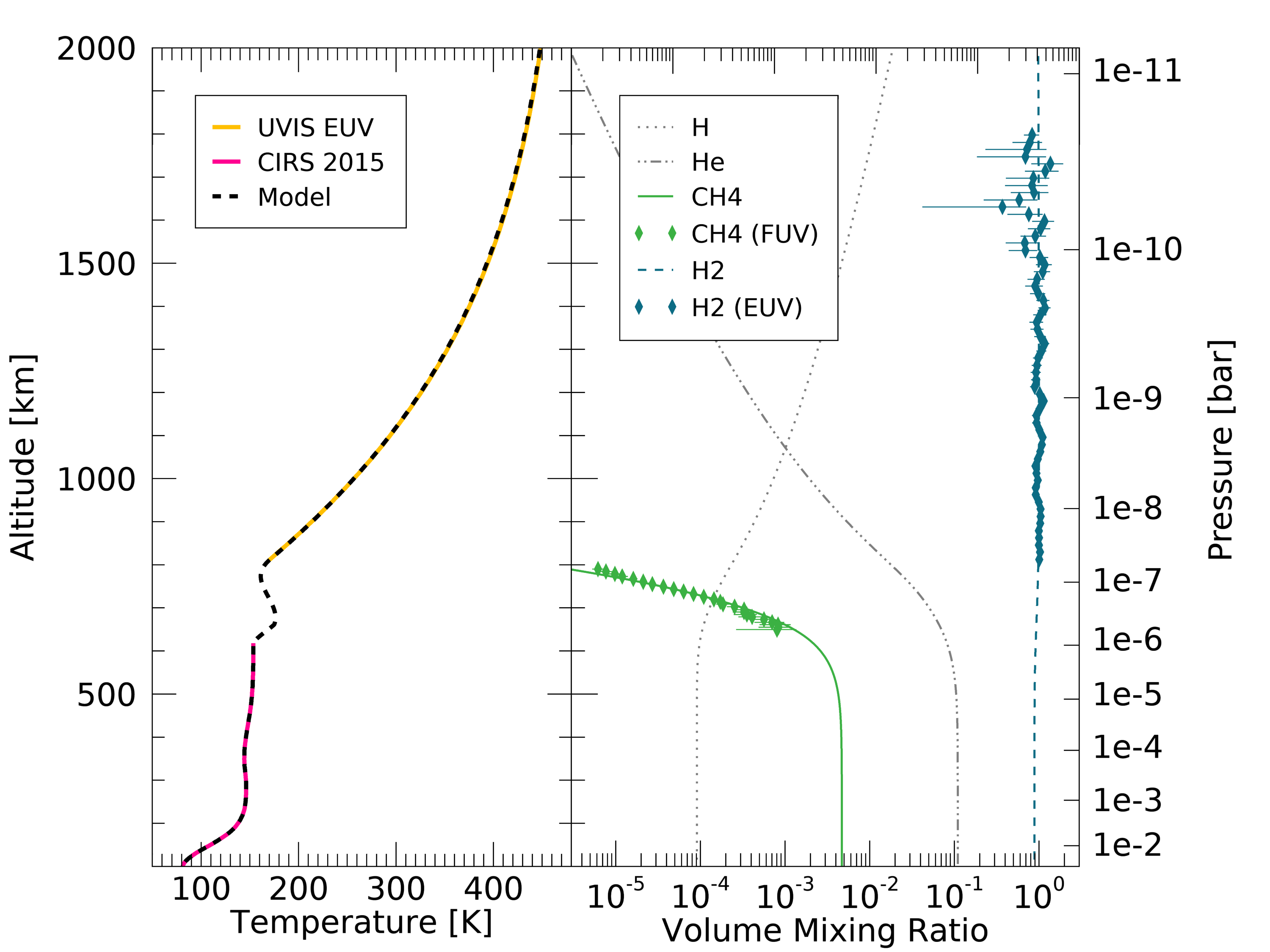}
\centering
\caption{Atmospheric structure model for an occultation of Zeta Orionis that probed the atmosphere at 42$^{\circ}$ N planetocentric latitude (FUV2017\textunderscore183\textunderscore00\textunderscore47\textunderscore29 in Table~\ref{table:UVISdata}). Pressures shown at right and altitudes above the 1 bar level shown at left are for both panels. Left: Model atmospheric temperature profile. Temperatures in the thermosphere retrieved from stellar occultations \citep{Brown20} are shown in yellow and temperatures retrieved from CIRS limb scans observed in 2015 in the stratosphere \citep{KG18} are shown in red. In the vertical region not covered by these two observations, temperatures are modeled by a spline interpolation with a pivot pressure to set a temperature maximum in order to produce a fit to the observed H\textsubscript{2} densities in the thermosphere (see text). Right: Model volume mixing ratios are shown for four key neutral species: H\textsubscript{2}, He, CH$_4$, and H. Blue and green diamonds show the VMRs based on observations by the UVIS EUV (H\textsubscript{2}) and FUV (CH$_4$) channels, respectively, using the atmospheric structure model as a total density reference.}
\label{fig:atmo}
\end{figure}

\noindent
We constructed atmospheric structure models for each occultation location to establish a pressure-altitude reference for our results by using the procedure established by \citet{KG18}. These models require a profile of temperature, which is not available in the coverage gap between CIRS and UVIS in the mesosphere. To address this, we construct a model temperature profile based on observations of the thermosphere and stratosphere. For each occultation latitude, we combined the CIRS temperature retrievals in the lower stratosphere and UVIS temperature profiles in the thermosphere. Specifically, we use the forward-model temperature profiles determined by \citet{Brown20}, fitted to the EUV occultation data. We interpolated the stratospheric temperatures retrieved from the CIRS data to the latitude of the UVIS occultations. We used one of three methods to fill in the temperatures for the pressure coverage gap between the CIRS and UVIS observations with the requirement that the model density profile matches the observed H\textsubscript{2} profile in the thermosphere \citep{KG18}. The method selected in each case depended on the best fit to the H\textsubscript{2} density profiles in the thermosphere from \citet{Brown20}. The first method used a simple cubic spline interpolation and the second a spline interpolation with a pivot pressure to set a temperature minimum or maximum in the gap region when necessary. The third method modeled the temperature profile with a classic Bates profile in the thermosphere, similar to that used by \citet{Brown20} to fit the UVIS EUV occultations. In some cases, a good fit to H$_2$ requires a local minimum or maximum in the atmospheric model temperature profile, as shown in Figure \ref{fig:atmo}. The presence of a maximum in the gap region should not be taken as a literal representation of the atmospheric conditions. In some cases, such a profile is necessary to fit the observed H$_2$ profiles. This need can arise due to both variability and uncertainty with the CIRS temperature profile as well as uncertainties in the shape of the 100 mbar surface, which are expected to be on the order of 15 km. A further discussion of the connection of these uncertainties to the overlying atmospheric structure in similar models can be found in \cite{KG18}. The creation of the atmospheric temperature profile allowed us to join the observed CIRS and UVIS temperatures smoothly and create a full atmosphere profile of temperature at each of our occultation latitudes. Our photochemical models also use these temperature profiles.

Based on the pressure-temperature profiles obtained above, we calculate the altitudes above the 0.1-bar level assuming hydrostatic equilibrium. This requires knowledge of the mean molecular weight profile that primarily depends on the relative abundances of H\textsubscript{2}, He, CH$_4$, and H where H\textsubscript{2} and He are most important. Having calculated trial altitudes based on constant mixing ratios for these species, we use equation (10.63) from \citet{schunk_ionospheres_2000} for partial pressures and thus volume mixing ratios of these species in diffusive equilibrium. Here, we fixed the volume mixing ratio of He below 100 mbar to 0.11 \citep{KG18} and of CH$_4$ to 4.7~$\times$~10$^{-3}$ \citep{FLETCHER09}. Then we recalculate the altitudes of the pressure levels and iterate to convergence. During this procedure, we ensure that our diffusive model profile for H agrees with the solar occultation upper limit of about 5\% at the top of the thermosphere \citep{KOSKINEN13}. The result depends on the molecular diffusion coefficients and the eddy diffusion (K$_{zz}$) profile. Mutual diffusion coefficients for He-H, He-H\textsubscript{2}, H-H\textsubscript{2}, H\textsubscript{2}-CH$_4$, and He-CH$_4$ pairs are given by \citet{Morero_Mason_72} and for the H-CH$_4$ pair by \citet{banks_aeronomy_1973}. We varied K$_{zz}$ with a profile that was either constant with pressure or followed a parameterized profile using equation (4) from \citet{KG18} in order to fit the methane profile observed by UVIS. Figure~\ref{fig:atmo} shows the atmospheric structure model for an occultation of Zeta Orionis that probed the atmosphere at 42.5$^{\circ}$ N planetocentric latitude. With a structure model for each occultation location, we can calculate volume mixing ratios (VMRs) and compare the observed hydrocarbon abundances to predictions by our seasonal photochemical model and the VMRs retrieved from the CIRS data. 

\subsection{Photochemical Model}
\label{M-JModel}

\noindent
We investigate the chemistry of Saturn's atmosphere at northern summer solstice by comparing the results of a photochemical model with the observed hydrocarbon abundances. We employ the Caltech/JPL KINETICS code \citep{Allen81,Yung84} to solve the time-dependent 1D continuity equations, employing orbital properties, ring shadowing, and incident solar flux that varies with season \citep[e.g., see][]{MG05}, as well as an updated chemical-reaction list \citep{MOSES23}. We start with fixed conditions for the equinox ($L_s = 0^{\circ}$) until steady state is reached, after which time-dependent conditions are introduced to incorporate seasonally changing solar insolation. The model produces vertical profiles spanning from the troposphere to the thermosphere for hundreds of species containing C, H, N, and O, including each of the five hydrocarbons observed by UVIS. The absorption cross sections and reaction rate coefficients for the neutral hydrocarbon and oxygen species are from \citet{MP17} and \citet{Moses18}. For the neutral nitrogen species, they are predominantly from previous exoplanet and Titan models \citep[e.g.,][]{Moses11,Moses13,Vuitton12,Vuitton19,Loison15}.

The Neutral Model includes 164 species and about 1100 reactions on a grid containing 260 pressure levels, ranging from 5 bar to 10$^{-8} \mu$bar. The Ion Model uses the same pressure grid, including solar-driven coupled ion-neutral photochemistry with a total of 321 species and over 2,500 chemical reactions. The ion reactions are from various sources, including models of interstellar chemistry \citep{Loison17}, dissociative recombination reviews \citep{Florescumitchell06} but primarily from Titan work, including laboratory work \citep{McEwan07} and models \citep{Vuitton19}. We updated the rate coefficients for reactions involving the \textsuperscript{1}CH\textsubscript{2} radical, using results from \citet{Douglas18}. This improved the agreement between the model and the observations whereas the previous rate coefficients from \citet{MP17} led our model to underestimate the abundance of C$_2$H$_2$ at all latitudes. The models also include condensation of water, iso-butane, and n-butane (C\textsubscript{4}H\textsubscript{10}) as well as diacetylene (C\textsubscript{4}H\textsubscript{2}) in Saturn's lower stratosphere.

As is typical for 1D photochemical models, transport was assumed to occur solely by vertical eddy and molecular diffusion, as described below. Mutual diffusion coefficients for H, He, CH$_4$, CO and CO\textsubscript{2} with H\textsubscript{2} come from laboratory data \citep{Morero_Mason_72}. Unfortunately, apart from CH$_4$, laboratory measurements of hydrocarbon molecular diffusion coefficients in H\textsubscript{2} atmospheres are not currently available. Previous models of Saturn's upper atmosphere relied on a simplistic approach, linking the C\textsubscript{x}H\textsubscript{y} molecular diffusion coefficients to that of CH$_4$ \citep{Moses00,MG05,Moses18}. However, the current analysis of UVIS data revealed that these assumptions were inadequate, highlighting the need for improved diffusion coefficients. To address this, we conducted theoretical calculations of the C\textsubscript{2}H\textsubscript{x} diffusion coefficients using Lennard-Jones potentials \citep{Reid87}. The molecular diffusion coefficients for various species are determined using the following equations:

\begin{equation}
C_2H_2: D = (5.26 \times 10^{17} \times T\textsuperscript{0.5})/n
\end{equation}
\begin{equation}
C_2H_4: D = (5.09 \times 10^{17} \times T\textsuperscript{0.5})/n
\end{equation}
\begin{equation}
C_2H_6: D = (4.74 \times 10^{17} \times T\textsuperscript{0.5})/n
\end{equation}

\noindent where, D represents the molecular diffusion coefficient in cm\textsuperscript{2}/s, T is the temperature in Kelvin, and n is the atmospheric number density in cm\textsuperscript{-3}, all of which vary with altitude. Nevertheless, there remains a need for new laboratory measurements of H\textsubscript{2}-C\textsubscript{x}H\textsubscript{y} molecular diffusion coefficients, particularly for C$_2$H$_2$, C$_2$H$_4$, C$_2$H$_6$, and C$_6$H$_6$, to facilitate more accurate comparisons with Cassini observations.

Note that uncertainties in chemical rate coefficients, photolysis cross sections, diffusion coefficients, and other model input parameters will translate to uncertainties in the final constituent mixing ratios predicted by the model. Although we do not formally or explicitly track model uncertainties in this paper, we note that previous photochemical models that have included rate-coefficient uncertainty propagation via Monte Carlo techniques for Saturn \citep[e.g.,][]{Dobrijevic03}) suggest that model uncertainty factors are less than 2 for C$_2$H$_2$ and C$_2$H$_6$ and less than 3 for C$_2$H$_4$ but typically increase with increasing molecular weight, becoming as much as an order of magnitude for species with four or more carbon atoms.  Although benzene was not included in the \citet{Dobrijevic03} study, similar investigations that include ion chemistry for Titan and Neptune suggest that model-predicted mixing ratios for benzene have significant uncertainties \citep[e.g.,][]{Loison19,Dobrijevic03}.

A set of 1D time-variable seasonal models for different latitudes concurrent with some of the UVIS observations were run. The Neutral Model covers 19 occultation latitudes, spanning from 85.7$^{\circ}$N planetocentric latitude to 85.7$^{\circ}$S planetocentric latitude. Initial conditions were set to a fully converged, fixed-season, steady-state solution at L\textsubscript{s}=0 (northern spring equinox). Subsequently, the model evolved with time, considering incident solar flux variations due to seasonally varying planetary geometry, orbital position, and ring shadowing relevant to each latitude \citep[see][for further details]{MG05}.  We assumed a constant solar-cycle average flux for the calculations and ran the neutral model for multiple Saturn years, repeating conditions relevant to the current Saturn year from $L_s = 0^{\circ}$ in August 2009 to $L_s = 360^{\circ}$ in January 2039. To account for seasonal changes in insolation, we updated the incident solar flux every two days in accordance with Saturn's changing heliocentric distance and seasonally changing axial tilt. The effect of the ring shadow is included in plane-parallel geometry where the rings are divided into five broad regions with averaged optical depths derived from Voyager data \citep{Sandel82, Holberg82}; see \citet{MG05} for further details. The model includes an isotropic Lyman $\alpha$ source with a brightness of 210 R from the scattering of solar Lyman $\alpha$ by hydrogen in the interplanetary medium \citep[e.g.,][]{Koskinen20}. The Neutral Model is run in time-variable seasonal mode for six Saturn years, with year 6 showing no notable differences compared to year 5.

The K$_{zz}$ profile in the photochemical model was adjusted at each latitude to achieve a reasonable agreement with both the CH$_4$ retrievals from the UVIS data as well as the C$_2$H$_6$ and C$_2$H$_2$ retrievals from Cassini CIRS limb and nadir observations \citep[e.g.,][]{Guerlet15,FLETCHER15,fletcherbook18,Fletcher20} (the functional form of the eddy diffusion coefficient, K$_{zz}$, is given in Appendix A). Initially, we adjusted the K$_{zz}$ profiles to match the CH$_4$ profiles retrieved from the UVIS data and assumed a constant K$_{zz}$ with latitude at deeper pressure levels in the stratosphere. This approach, however, failed to match the meridional distributions of C$_2$H$_6$ and C$_2$H$_2$ in the stratosphere (see Figure \ref{fig:DiffKzz}). Subsequently, we developed an alternative K$_{zz}$ distribution with a latitude-dependent K$_{zz}$ in the lower stratosphere where the value of K$_{zz}$ varies by about an order of magnitude from equator to poles, with weaker mixing at the poles. For example, at the 1-mbar level our $K_{zz}$ coefficient varies from about 6 m$^2$~s$^{-1}$ around the equator to 0.2--1 m$^2$~s$^{-1}$ at high latitudes. This revised K$_{zz}$ profile resulted in an improved agreement of the model with the CIRS data for C$_2$H$_2$ and C$_2$H$_6$, and we adopted it for both the Neutral and Ion Models. We note that the change in K$_{zz}$ values in the stratosphere does not substantially affect the pressure-altitude relation in the atmospheric structure models or our determination of the homopause pressure.

The Ion Model was run on the same pressure grid and at the same latitudes as the Neutral Model, with the exception of the highest planetocentric latitudes of 81.4$^{\circ}$ N and 85.7$^{\circ}$ N, where the model running time increased significantly. This is because the large seasonal changes to insolation at high latitudes lead to much longer convergence times, compared to the equatorial regions where the overhead solar illumination is more consistent. The Ion Model run (using the revised K$_{zz}$ profile described above) lasted approximately three-quarters of a Saturn year, starting from L\textsubscript{s} = 0$^{\circ}$ and extended to the fall equinox, with specific outputs used in this work for L\textsubscript{s} = 90$^{\circ}$, corresponding to the northern summer solstice. The inclusion of ion chemistry generally led to unfeasible running time even at low latitudes. Due to computational limitations, it was therefore not possible to run longer simulations with the Ion Model. Despite the truncated run, the results in the upper stratosphere and mesosphere can still be compared with the UVIS data because chemical time constants for the hydrocarbons in the homopause region where the observations are sensitive are much shorter than a Saturn season. As a result, it is expected that the abundances would not significantly differ during subsequent years of the simulation. One exception is for CH$_4$ and C$_2$H$_6$ at high latitudes in the winter hemisphere, where chemical loss time scales approach a Saturn season. Because the chemical time constants become large in the middle stratosphere, however, the results for this region are still influenced by the initial conditions, which were for L\textsubscript{s} = 0 $^{\circ}$, and direct comparisons of the Ion Model results with the CIRS data should be avoided. In this way, the Neutral Model comprises a robust framework for comparison to both the CIRS and UVIS datasets and the Ion Model allows for inferences to be made about the impact of ion chemistry on hydrocarbon abundances in the upper stratosphere and mesosphere. An improved ion chemistry model that also includes auroral particle precipitation will be pursued in future work.

\section{Results}
\label{Results}

\noindent
In this section, we present the findings from the UVIS and CIRS retrievals and compare them to the photochemical model results, beginning with the UVIS hydrocarbon retrievals. First, we describe the distribution of CH$_4$ and locate the homopause at different latitudes. We then investigate the 2D distributions of C$_2$H$_6$, C$_2$H$_4$, and C$_2$H$_2$ retrieved from the UVIS data, highlighting a clear seasonal trend in their distribution at northern summer solstice. We also analyze vertical profiles of all observed species near 30$^{\circ}$, 60$^{\circ}$, and 70--75$^{\circ}$ latitude in both hemispheres to compare regions with different photochemical conditions, including the low latitudes near the subsolar point, middle latitudes affected by shadowing from the rings, and auroral regions with unique auroral processes. Furthermore, we compare the UVIS and CIRS observations with the predictions of the photochemical model. Lastly, we analyze the distribution of C$_6$H$_6$, which is expected to be significantly affected by ion chemistry, and compare it with the photochemical model predictions.

\begin{figure}
\noindent\includegraphics[width=\textwidth]{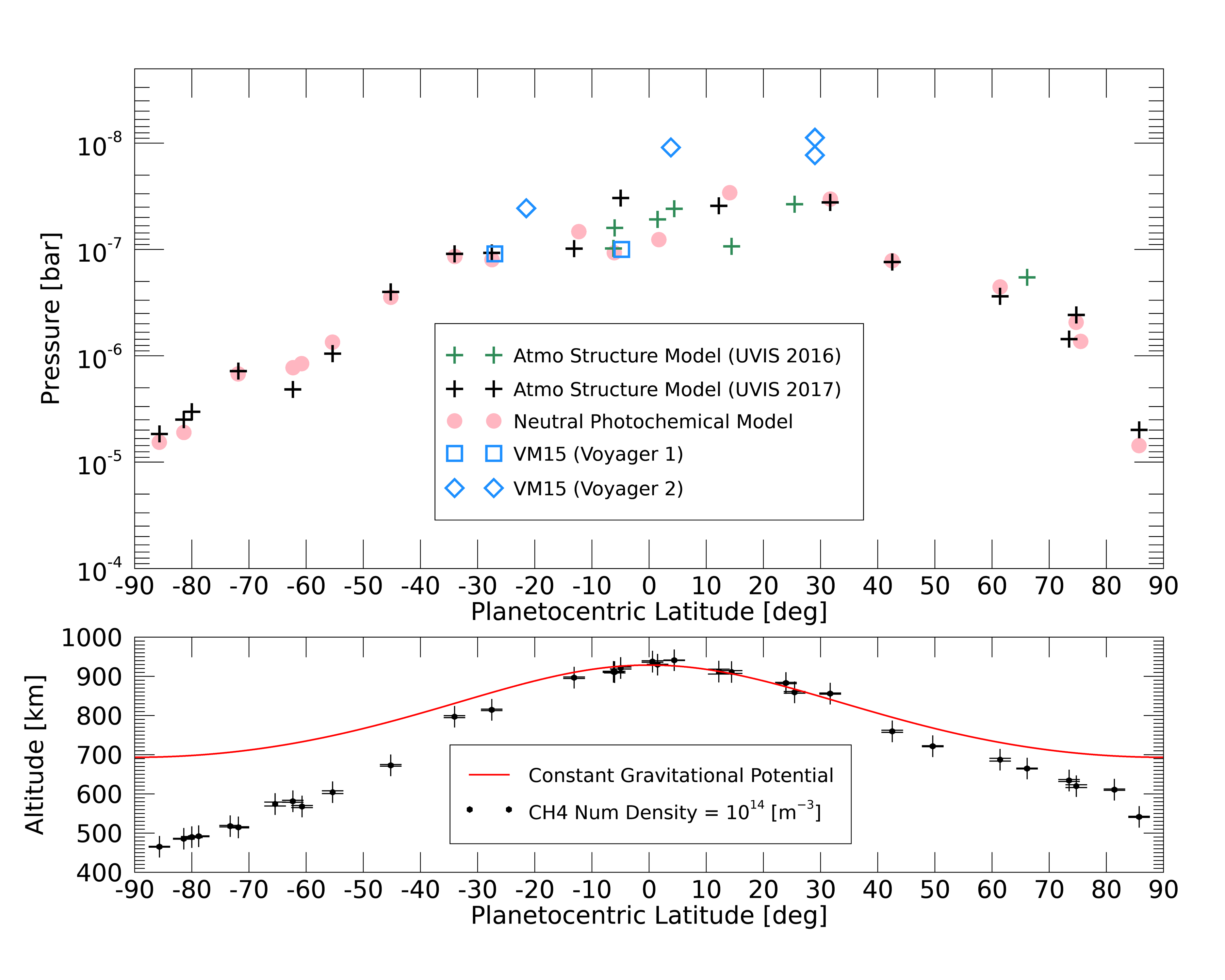}
\caption{Trends in the homopause level as a function of latitude. The upper panel shows the homopause reference level as a function of latitude from the atmospheric structure model (plusses, 2016 in green and 2017 in black) and from the photochemical model (pink circles), defined here as where the CH$_4$ volume mixing ratio equals 10$^{-5}$. We also show the corresponding homopause pressures retrieved from Voyager solar and stellar occultations by \citet{VM15} in blue (boxes for Voyager 1 and diamonds for Voyager 2). The lower panel shows the altitude where the density of methane equals 10\textsuperscript{14} m\textsuperscript{-3} (plusses). A constant gravitational surface potential level corresponding to an equatorial altitude of 927 km above the 1-bar level is shown in red. The differences in altitude exceeds 150 km at the north pole and 200 km at the south pole.}
\label{fig:Homo Trend}
\end{figure}

\subsection{Methane: An unexpected meridional trend in homopause depth}
\label{R-Homo}

\noindent
As the most abundant hydrocarbon, the photochemical lifetime of CH$_4$ is sufficiently long for its distribution to be primarily controlled by dynamical mixing. In particular, the CH$_4$ profiles can be used as a tracer of the homopause. The homopause marks the altitude or pressure level below which the atmosphere is dominated by eddy diffusion and is therefore well-mixed and above which molecular diffusion dominates and the abundance of species with height is dictated by their individual mass-dependent scale heights. The abundance of CH$_4$ decreases rapidly with altitude above the homopause. While dynamical mixing is represented by the eddy diffusion coefficient in 1D photochemical models, the pressure of the homopause is also affected by global circulation \citep[e.g.,][]{Mueller-Wodarg03}. For example, upwelling moves the homopause to lower pressures (higher altitudes) by transporting methane-rich air from below. Conversely, downwelling brings methane-poor air down, leading to a higher homopause pressure (at lower altitudes). 

Our analysis of the Grand Finale UVIS occultations confirms earlier indications that the depth of the CH$_4$ homopause is deeper at the poles than at the equator \citep[e.g.,][]{Gerard09}. The extent of the difference that we observe, however, is large and systematic with latitude. The upper panel of Figure~\ref{fig:Homo Trend} shows the homopause reference level (see Section~\ref{sc:intro}) from our atmospheric structure models (see Section \ref{M-Atmo}) and the neutral photochemical model (see Section \ref{M-JModel}). The reference level increases by about two decades in pressure from the subsolar latitudes to the poles. There is a relatively good agreement between our results and the Voyager 1 observations, while the Voyager 2 retrievals indicate a somewhat lower-pressure homopause around the equator. We note that more subtle changes in the homopause level can be found on smaller spatial scales, with, for example, somewhat lower homopause pressures near 70$^{\circ}$N and 70$^{\circ}$S. While CH$_4$ cannot be observed in the UVIS data to the same depth at all latitudes, the variation in CH$_4$ density with latitude extends down to about the 10 $\mu$bar level, below which the density contours flatten out and become approximately constant with latitude (see Figure \ref{fig:VMR}).

We note that this homopause trend cannot be caused by Saturn's oblate shape. The observed trend is significantly greater than the expected variation in density with latitude based on the oblate shape. To demonstrate this, the lower panel of Figure \ref{fig:Homo Trend} compares the altitude above the 1 bar level of a CH$_4$ constant density surface that is close to the homopause with a surface of constant gravitational potential. The altitude of the constant density surface is up to 150 km deeper than predicted by the surface of constant gravitational in the northern hemisphere and up to 220 km deeper in the southern hemisphere. We note that the shape of the atmosphere at the 100 mbar level based on the \citet{AS07} gravitational potential that we use agrees reasonably well with radio occultations and has an uncertainty of about $\pm$15 km \citep[e.g.,][]{KG18}. The shape at the lower pressures considered here depends largely on the underlying temperature structure. The variation in altitudes of the pressure levels at the top of the mesosphere, however, is only about 30 km and this includes seasonal changes in temperatures as observed by CIRS \citep{Koskinen15}. Thus, changes in temperature in the stratosphere and mesosphere cannot explain the observed trend. Finally, this trend is not driven by dissociation or photochemistry. The homopause is found at lower pressures around the subsolar point and higher pressures at high latitudes, in contrast to the expectations based on seasonal insolation. Also, our photochemical model obviously includes photolysis of CH$_4$ and still requires a latitude-dependent K$_{zz}$ profile to accurately reproduce the observed CH$_4$ distribution. 

\begin{figure}
\noindent\includegraphics[width=\textwidth]{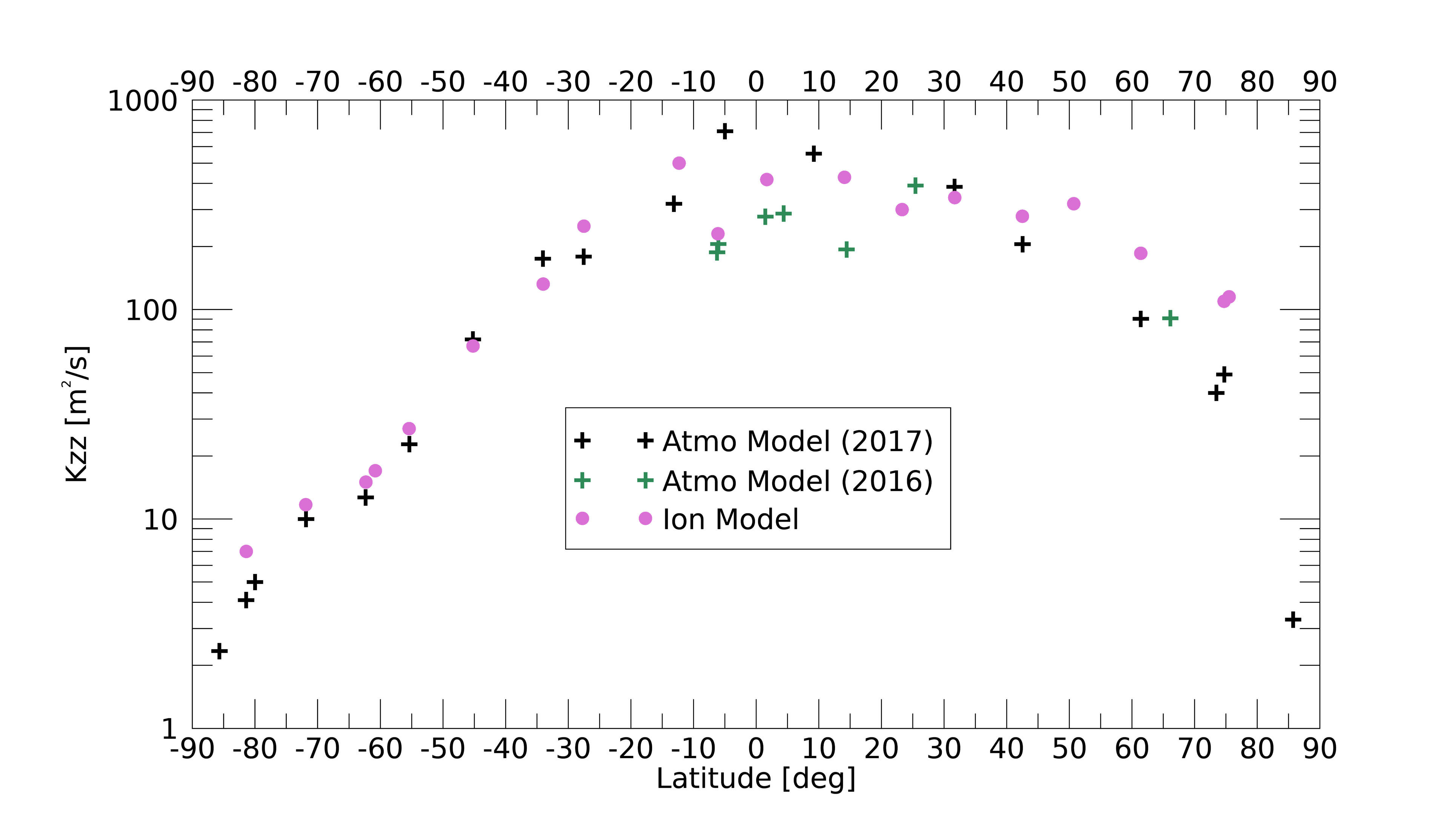}
\caption{The modeled K$_{zz}$ at 2 $\mu$bar. Fits from the atmospheric structure model are represented by plusses (models for observations from 2017, encompassing the Grand Finale tour, are show in black and those from 2016 in green). Fits to the data from the ion-neutral photochemical model are represented by pink circles.}
\label{fig:Kzz}
\end{figure}

Thus, the observed homopause trend indicates that mixing is weaker at high latitudes than around the sub-solar point. Figure \ref{fig:Kzz} illustrates this by showing K$_{zz}$ from our models as a function of latitude at the pressure of 2 $\mu$bar i.e., near the homopause at high latitudes. We note that the atmospheric structure model and photochemical model K$_{zz}$ values are not identical, but are reasonably similar around the homopause level (see Figure \ref{fig:Kzz}). In many applications, K$_{zz}$ is assumed to represent mixing by turbulence and breaking atmospheric waves. Our results could indicate that turbulence and/or wave propagation is suppressed around the poles. Alternatively and perhaps more likely, the homopause trend could be due to global circulation. This would imply that at high latitudes, there is a transport of air from higher altitudes, where CH$_4$ concentrations are lower, towards lower altitudes, resulting in a decrease in CH$_4$ abundance with altitude. Correspondingly, we might expect there to be upwelling near the subsolar point that would have an opposite effect. Intriguingly, the 2016 occultations are consistent with slightly lower values of K$_{zz}$ around the equator than the Grand Finale occultations. This could indicate that upwelling around the equator strengthened towards the solstice in 2017. We discuss these possibilities further in Section \ref{D-Homo}.

\subsection{Meridional and vertical profiles of the photochemical products}
\label{R-Hydro}

\begin{figure}
\noindent\includegraphics[width=\textwidth]{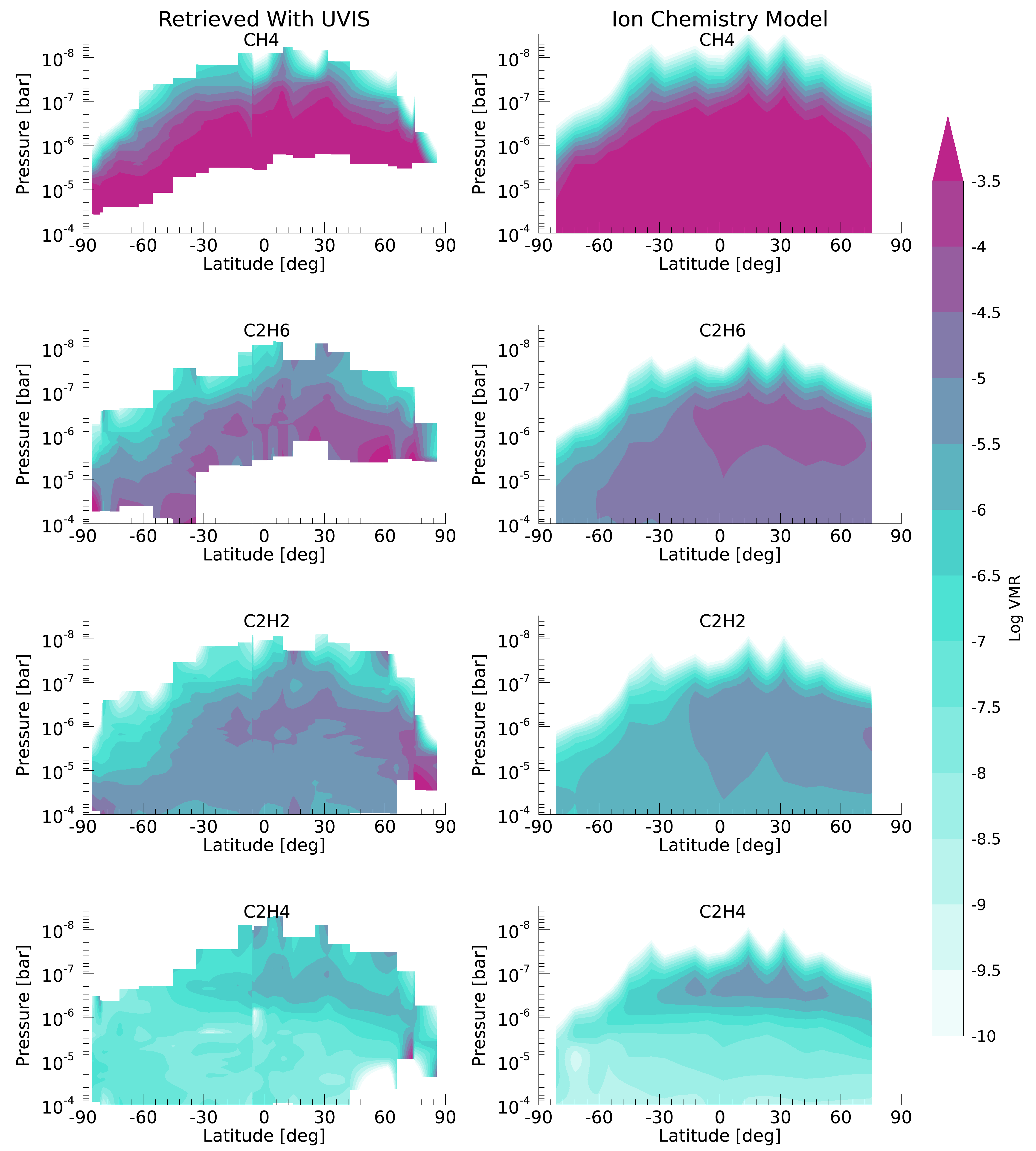}
\caption{Volume mixing ratios interpolated over latitude and pressure for CH$_4$, C$_2$H$_2$, C$_2$H$_4$ and C$_2$H$_6$ observed by UVIS (left) and predicted by the Ion Model (right). The meridional trend in the CH$_4$ homopause discussed in Section~\ref{D-Homo} is observable in the CH$_4$ distribution (upper left) and reflected by the low-pressure limit for detectability with latitude for the other species. Along isobars, abundances generally peak at low latitudes in the northern (summer) hemisphere near the subsolar point as predicted by the seasonal photochemical model.}
\label{fig:VMR}
\end{figure}

\noindent
The meridional trend in the homopause reference level is clearly reflected in the distributions of the photochemical products. This is illustrated by Figure \ref{fig:VMR}, which shows the 2D distribution of volume mixing ratios for CH$_4$, C$_2$H$_6$, C$_2$H$_4$ and C$_2$H$_2$ as a function of latitude and pressure based on the UVIS observations. Note that we did not detect C$_6$H$_6$ at southern middle latitudes, and hence it is not shown in the figure. In broad agreement with the photochemical model, the abundances of C$_2$H$_2$, C$_2$H$_4$ and C$_2$H$_6$ below the homopause are generally higher at low latitudes and in the northern (summer) hemisphere. Interestingly, the abundances peak at slightly lower pressures/higher altitudes than at surrounding latitudes near 30$^{\circ}$ N planetocentric latitude (close to the subsolar point at 26.9$^{\circ}$ N planetocentric latitude). There is also a local maximum in the abundances of C$_2$H$_6$, C$_2$H$_2$, and C$_2$H$_4$ at 75$^{\circ}$ N at about 10 $\mu$bar. This maximum roughly coincides with auroral latitudes \citep{Lamy18} where hydrocarbon production could be enhanced by electron and ion precipitation \citep{Gerard09,Galand11}. It is not predicted by our photochemical models because the models do not include these processes. In the winter hemisphere, there is a local minimum in the abundances of C$_2$H$_6$ and C$_2$H$_2$ centered around 60$^{\circ}$S planetocentric latitude at pressures spanning $\sim$1 to $\sim$10 $\mu$bar that is not seen in either of the photochemical models. It is possible that downwelling in this ring-shadowed region could be playing a role.

\begin{figure}
\noindent\includegraphics[width=\textwidth]{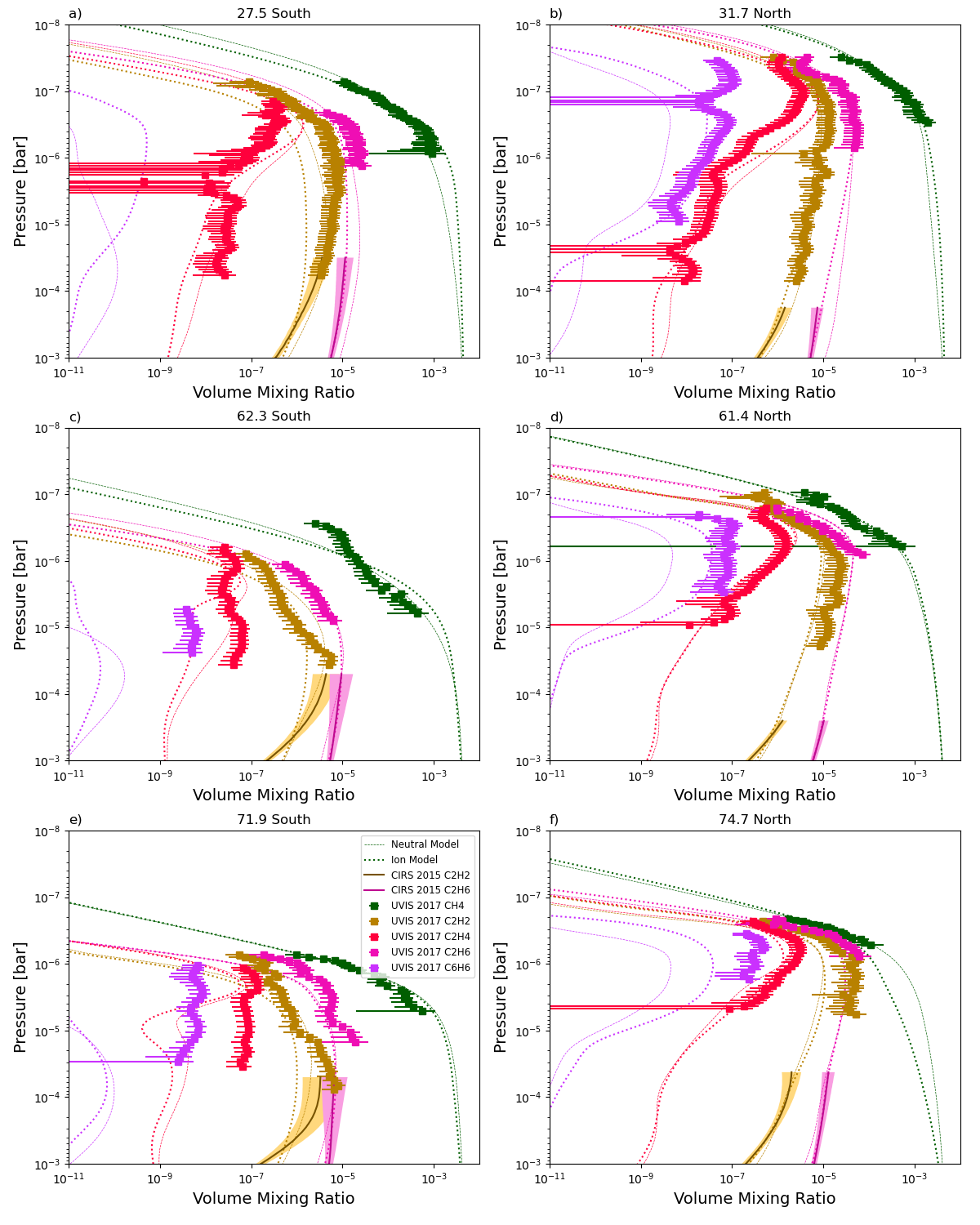}
\caption{Volume mixing ratios as a function of pressure for CH$_4$, C$_2$H$_6$, C$_2$H$_4$, C$_2$H$_2$ and  C$_6$H$_6$ at northern (summer, right) and southern (winter, left) latitudes (see legend in panel c; note: the subsolar point for the Grand Finale period is located at 26.9$^{\circ}$ N planetocentric latitude). The mixing ratio profiles of C$_2$H$_2$ and  C$_6$H$_6$ retrieved from CIRS limb scans are shown by the solid lines inside the shaded regions. Predictions from the photochemical models are overplotted with the Neutral Model results shown by the dashed lines and the Ion Model results shown by the dotted lines.}
\label{fig:ndensprofile}
\end{figure}

\begin{figure}
\noindent\includegraphics[width=\textwidth]{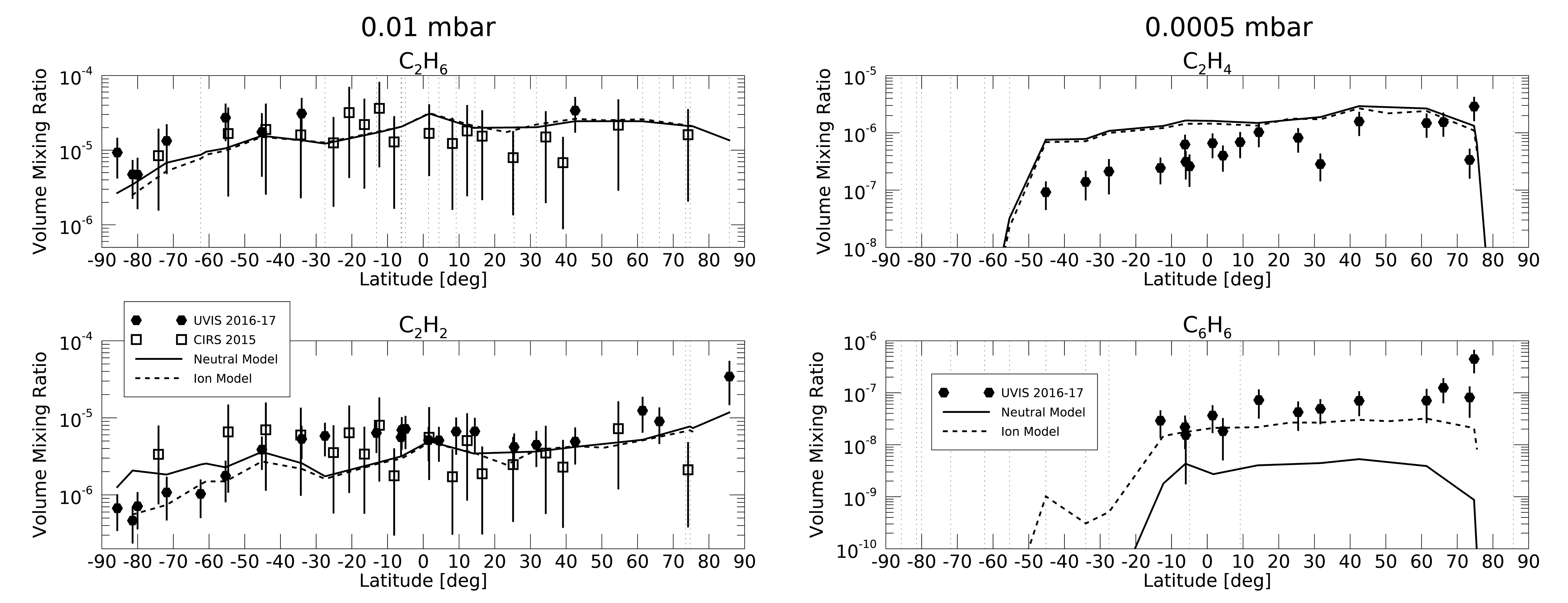}
\caption{Meridional trends in hydrocarbon volume mixing ratio with latitude. The left hand side shows C$_2$H$_6$ and C$_2$H$_2$ at 0.01 mbar and the right shows C$_2$H$_4$ and C$_6$H$_6$ at 0.0005 mbar. Data collected by UVIS during 2016-2017 are represented by black hexagons (with a vertical grey dotted line indicating latitudes beyond UVIS's measurement range at this pressure level), while points observed by CIRS in 2015 are denoted by squares. These are compared to outputs from photochemical models, with the solid line representing the Neutral model and the dashed line representing the Ion model.}
\label{fig:meridionaltrends}
\end{figure}

Vertical profiles of the hydrocarbons constrain photochemistry and dynamics at different pressure levels and facilitate a deeper analysis of the results. Combining the UVIS results with the profiles retrieved from CIRS significantly extends the scope of these constraints. In Figure~\ref{fig:ndensprofile}, we compare the UVIS hydrocarbon mixing ratio profiles with CIRS profiles for C$_2$H$_2$ and C$_2$H$_6$ as well as the results from our two photochemical models. For this comparison, we chose representative latitudes near 30$^{\circ}$, 60$^{\circ}$ and 70-75$^{\circ}$ planetocentric latitudes in both hemispheres. Starting with northern middle latitudes, we examine the profiles at 31.7$^{\circ}$N. Here, the observed profiles of C$_2$H$_4$ and C$_2$H$_2$ are reproduced reasonably well by both the Ion and Neutral Models. The C$_2$H$_4$ profile is well-fit below the 2 $\mu$bar level, but the models overpredict the peak abundances at lower pressures. The C$_6$H$_6$ is better fit by the Ion Model and the agreement is good below the 1 $\mu$bar level, however, at lower pressures the observed C$_6$H$_6$ mixing ratios exceed the model predictions. As expected, the Neutral Model predicts C$_6$H$_6$ abundances that are much lower than those predicted by the Ion Model \citep{Koskinen16}. The CIRS observations of C$_2$H$_2$ align well with both the model predictions and the UVIS observations at this latitude. This level of agreement between the instruments is remarkable, considering the different methods that are used to retrieve hydrocarbon abundances from the UVIS and CIRS observations, and provides a confirmation of both approaches. 

Moving farther north to 61.4$^{\circ}$N, we find that the C$_2$H$_6$ abundance predicted by the photochemical models is too high by a factor of 1.2 to 3.2 few at pressures lower than 7~$\times$~10$^{-7}$. Both the Ion and Neutral Models agree well with the observed C$_2$H$_2$ abundances between 2~$\times$~10$^{-7}$ and 5~$\times$~10$^{-7}$ bar, near the homopause. However, at pressures higher than 6~$\times$~10$^{-7}$ bar, the models underestimate the observed abundances by a factor of 1.5 to 3. The models achieve a relatively good fit to the C$_2$H$_4$ profile at this latitude, although they tend to slightly overpredict the peak abundance. The C$_6$H$_6$ profile is better fit by the Ion Model than the Neutral Model, although the observed profile is more constant with pressure than predicted by the model and exceeds the Ion Model abundances by up to a factor of about five. 

At southern middle latitudes, in the region shadowed by the rings, we begin by examining the results for 27$^{\circ}$S planetocentric latitude. Here, the observed abundances of C$_2$H$_2$ and C$_2$H$_6$ are larger than predicted by the models. Specifically, we observe a C$_2$H$_2$ profile that more closely follows the Neutral Model but which still exceeds the predicted abundances above the 10 $\mu$bar pressure level, especially around the peak at 0.2 $\mu$bar where the observed mixing ratios are higher by a factor of 5 than the Neutral Model prediction. Similarly, the C$_2$H$_6$ profile is better fitted by the Neutral Model but the observed mixing ratios still exceed the model prediction by approximately a factor of two. As in the north, the abundance of C$_2$H$_4$ is lower than the model predictions by almost an order of magnitude at the pressures where the mixing ratio is expected to peak. Due to the reduced insolation arising from ring shadowing, the photochemical model predicts less C\textsubscript{2}H\textsubscript{X} at these latitudes. However, the observations do not indicate the same impact from ring shadow as the models suggest. This discrepancy may be attributed to either the rings being more light-permeable than the model parameterizes possibly due to the 'Venetian blind' effect (a phenomenon arising from ring dynamics, which leads to many small-scale regions of underdense material), or to the influence of dynamics on the abundances of these species. At pressures higher than 5 $\mu$bar, the observed C$_2$H$_4$ abundance follows the trend predicted by the Neutral Model reasonably well but exceeds the abundance predicted by the Ion Model. C$_6$H$_6$ is not detected at this latitude, indicating a lower mixing ratio than at similar latitudes in the north. Finally, the agreement between UVIS and CIRS C$_2$H$_2$ retrievals at this latitude is excellent, with the observations from these two instruments again providing more comprehensive vertical coverage. We note that differences in the CH$_4$ profile between the Ion and Neutral models arise from differences in the K$_{zz}$ profiles implemented for those two models.

At 62$^{\circ}$S, just inside the latitudes confined to the permanent night in the winter hemisphere, the CH$_4$ profile has an unusual shape that does not agree well with the model profiles. This is because the observed profile cannot be precisely fit with the functional form for the K$_{zz}$ profile that we have adopted, indicating that more complex dynamics is required to explain the observed CH$_4$ profile at this location. The C$_2$H$_2$ profile here exhibits a similar shape to the CH$_4$ profile, including a local minimum at a pressure level where the models predict a peak, again suggesting the presence of physical drivers not accounted for in the model. This is confirmed by the CIRS observations that agree exceptionally well with our retrieval at pressures where they overlap. The C$_2$H$_4$ profile is not well-matched by the photochemical models either. In contrast, the observed C$_2$H$_6$ profile shows good agreement with both the Ion and Neutral Models but this may be coincidental, given the disagreements of the models with the observed profiles for the other species. Given that enhanced mixing could flatten the abundance profiles with pressure, smoothing out the predicted peaks, it is tempting to assign the differences between the models and the observed C$_2$H$_2$ and C$_2$H$_4$ profiles to the presence of downwelling that could also explain the CH$_4$ distribution. C$_6$H$_6$ is observed at this latitude with mixing ratios that far exceed those predicted by either model, with a difference of approximately two orders of magnitude. There is no prospect for enhanced production of C$_6$H$_6$ at this latitude in the winter hemisphere that is also remote from auroral production \citep[approximately 70$^\circ$ to 80$^\circ$ latitude][]{Lamy18}. Meridional transport and downwelling could potentially also explain the C$_6$H$_6$ abundances.

Turning now to the auroral profiles, at 74.7$^{\circ}$N, the photochemical models accurately predict the C$_2$H$_6$ profile. The abundance of C$_2$H$_2$, however, is underpredicted by the models by almost an order of magnitude at all but the lowest observed pressures. The observed C$_2$H$_2$ mixing ratios are comparable to those of C$_2$H$_6$, indicating that the C$_2$H$_2$/C$_2$H$_6$ ratio is enhanced at auroral latitudes. The observed C$_2$H$_4$ mixing ratios roughly agree with the models. The observed C$_6$H$_6$ profile is better fit by the Ion Model, but the observed mixing ratios still exceed predicted values by up to an order of magnitude, which could be an indication of enhanced production in the auroral region. 

At 71.9$^{\circ}$S, the observed C$_2$H$_6$ profile is not well-fit by the models except near 6 $\mu$bar; the models predict abundances that are too low by up to a factor of five. The remaining species bear much stronger similarities to the profile at 62.3$^{\circ}$S, than to those at 74.7$^{\circ}$N. The C$_2$H$_2$ abundances increase with pressure down to the bottom of the profile where they exceed those measured by CIRS by a factor of a few (although within error bars at some points). Given the good agreement between UVIS and CIRS at other latitudes, the difference in this case could be due to temporal variability based on the 2-year time difference between the CIRS and UVIS observations. With similarities to 62.3$^{\circ}$S, the observed C$_2$H$_4$ profile is relatively constant with pressure and is not well-fit by the models. The C$_6$H$_6$ profile is also near-vertical and exceeds the abundances predicted by the models by two or more orders of magnitude. Again, the observed hydrocarbon abundance profiles being relatively constant with pressure could be the result of downwelling at high southern latitudes.

In addition to the vertical profiles and contour plots, it is useful to explore the meridional mixing ratio profiles at select pressure levels. This exercise, however, is limited by the fact that not all species are detectable at all pressures, especially because of the homopause trend at low pressures. On the left hand panels of Figure~\ref{fig:meridionaltrends}, we compare the UVIS, CIRS, and model results for C$_2$H$_2$ and C$_2$H$_6$ at 0.01 mbar where both UVIS and CIRS results are available. We see that the agreement between UVIS and CIRS results is relatively good, given the uncertainties and possible temporal variability between observations. Both species show higher abundances in the summer hemisphere, with a clearer trend for C$_2$H$_2$, as expected based on the models. On the right hand panels of the figure, we show meridional trends at 0.0005 mbar for C$_2$H$_4$ and C$_6$H$_6$, which are only routinely detected by UVIS. The mixing ratio of C$_2$H$_4$ decreases with latitude from the summer to the winter hemisphere, with a gradient that is not reproduced by the models in the winter hemisphere. The Ion Model matches the C$_6$H$_6$ abundances relatively well in the summer hemisphere, albeit falling short at high latitudes. The models predict undetectable abundances for C$_6$H$_6$ in the winter hemisphere at this pressure, an expectation confirmed by the observations.

Overall, the model predictions show varying degrees of agreement with the observations across different latitudes, with generally better agreement at northern (summer) middle latitudes than at corresponding latitudes in the southern (winter) hemisphere. This is perhaps not surprising, given that production peaks in the summer hemisphere whereas dynamics and other processes are more important in the winter hemisphere. The Ion and Neutral Models show relatively small differences with each other, especially in the northern hemisphere and at higher latitudes, with the notable exception of C$_6$H$_6$. As expected, the Ion Model comes much closer than the Neutral Model to reproducing the observed C$_6$H$_6$ abundances considering the more efficient production of polycyclic hydrocarbons by ion chemistry \citep[e.g.,][]{Koskinen16}. At auroral latitudes, the abundances of C$_2$H$_6$, C$_2$H$_2$, C$_2$H$_4$ and C$_6$H$_6$ all exceed the model predictions at nearly every pressure level. The auroral abundances also generally exceed those near 60$^{\circ}$ latitude in the corresponding hemisphere, suggesting that auroral chemistry is enhancing the production of these hydrocarbons. The difficulty in fitting the CH$_4$ profile at 62$^{\circ}$S with our functional form of K$_{zz}$, however, suggests that dynamics may also play an important role at this latitude and at southern auroral latitudes. Additionally, the peculiar abundance profiles of C$_2$H$_2$ and C$_2$H$_4$ at southern (winter) high latitudes (and to some extent the abundance profile of C$_6$H$_6$), where their abundances do not decrease with depth as expected, might be indicative of a middle to high latitude downwelling cell that would also be consistent with the meridional trend in the homopause reference level. 

\subsection{Benzene and Ion Chemistry}
\label{R-Benzene}

\noindent
We detected C$_6$H$_6$ in all the stellar occultations except three occultations observed in the southern hemisphere between 20$^{\circ}$S and 50$^{\circ}$S planetocentric latitude. These latitudes fall within the region of the ring shadow where solar insolation is either blocked or reduced. The vertical extent of the observable C$_6$H$_6$ profile at other latitudes exhibits some variations, with the observations probing higher pressures in the south (between 100 and 1 $\mu$bar) and a broader range in pressure in the north (between 40 and 0.01 $\mu$bar, varying with latitude), following the trend observed in the other hydrocarbons that is set by the CH$_4$ homopause variation with latitude. We find that the observed C$_6$H$_6$ abundances peak near 75$^{\circ}$ N planetocentric latitude between 1 and 0.1 $\mu$bar. Given the greater efficiency of ion chemistry and auroral processes in generating C$_6$H$_6$ and other polycyclic hydrocarbons \citep{Wong03, Vuitton08}, the presence of this peak within the auroral region is expected. However, it is worth noting that a minor contribution to production rates occurs during polar winter due to the scattering of Lyman $\alpha$ photons from the local interplanetary medium. An interesting result is the presence of C$_6$H$_6$ in the southern hemisphere between about 60$^{\circ}$S and 70$^{\circ}$S planetocentric latitude, outside of the auroral region between roughly 70$^{\circ}$S and 80$^{\circ}$S latitude. At the time of these observations, this region had been in permanent shadow for at least 624 Earth days, pointing to auroral processes as drivers of C$_6$H$_6$ chemistry at these latitudes.

\begin{figure}
\noindent\includegraphics[width=\textwidth]{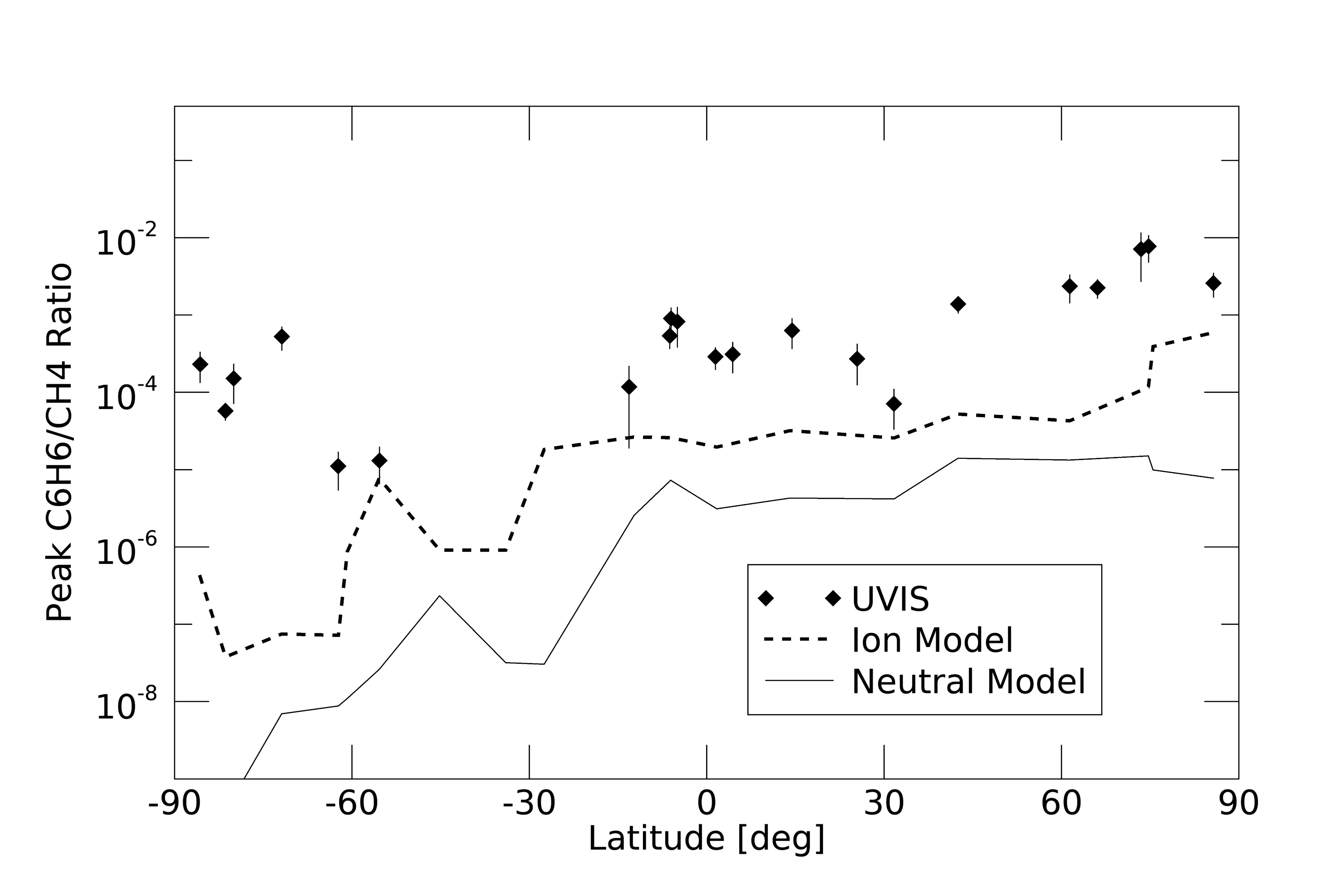}
\caption{Maximum C$_6$H$_6$ to CH$_4$ number density ratios between 0.1 and 10 $\mu$bar as a function of latitude following Figure 3 in \citet{Koskinen16}. These values exceed those predicted from the Neutral Model (dashed lines) and the Ion Model (dotted lines), though the Ion Model is closer.}
\label{fig:peakbenzene}
\end{figure}

The distribution of C$_6$H$_6$ in the upper atmosphere was previously characterized by \citet{Koskinen16} based on observations made between 2005 and 2015. They found that the peak C$_6$H$_6$/CH$_4$ volume mixing ratio increases with latitude in the northern hemisphere during spring (Figure 3 of that work). For comparison, we extract the peak C$_6$H$_6$/CH$_4$ volume mixing ratio from our observations and compare the observed ratios with the photochemical model results in Figure \ref{fig:peakbenzene}. In agreement with \citet{Koskinen16}, the peak C$_6$H$_6$/CH$_4$ ratio increases from the equator to the northern auroral region during the closely-following northern summer solstice time period. Notably, this increase in the peak C$_6$H$_6$/CH$_4$ volume mixing ratio appears to closely track the daily mean solar actinic flux input for the season. While the inclusion of solar-driven ion chemistry in our photochemical model enhances the predicted peak C$_6$H$_6$/CH$_4$ ratio compared to the Neutral Model, the Ion Model predictions remain too low at all latitudes except in the vicinity of approximately 60$^{\circ}$S. The disagreement between model and the observations is greatest at latitudes poleward of 60$^{\circ}$ S planetocentric latitude in the permanent night region of the winter hemisphere where we observe a large local maximum in the peak C$_6$H$_6$/CH$_4$ ratio. There is no solar insolation to drive the photolysis of CH$_4$ at these latitudes. Therefore, the results for C$_6$H$_6$ at southern high latitudes provide perhaps the strongest evidence for the influence of auroral processes that are not included in our photochemical models.

\section{Discussion}
\label{Discussion}

\noindent
In this section, we revisit several open questions raised in Section~\ref{Results} and discuss them in more detail. Maintaining a similar order, we start with an examination of the observed CH$_4$ distribution and explore potential mechanisms that could give rise to the meridional trend in the homopause reference level. Then, we investigate the behavior of the observed minor hydrocarbon distribution and further discuss the possible origin of the discrepancies between the observations and the photochemical models. Lastly, we focus on the high latitude results, with a particular focus on C$_6$H$_6$.

\subsection{Methane As a Tracer of Dynamics}
\label{D-Homo}

\noindent
In Section \ref{R-Homo}, we presented evidence of a significant variation in the homopause pressure level with latitude, where the pressure level corresponding to a constant CH$_4$ mixing ratio differs by two orders of magnitude between the equator and the poles. Effectively, the homopause trend leads the abundances of the minor hydrocarbons in the mesosphere to rapidly decrease with latitude away from the equator at pressures lower than about 10 $\mu$bar. At pressures higher than 10 $\mu$bar, this trend is no longer apparent and the CH$_4$ abundance becomes roughly constant with latitude. To accurately represent this meridional homopause trend, our photochemical model has been modified to incorporate a latitude-dependent K$_{zz}$ profile. This adjustment involves higher K$_{zz}$ values at low latitudes and lower values at high latitudes, especially in the southern hemisphere, similar to the atmospheric structure model K$_{zz}$ values depicted in Figure \ref{fig:DiffKzz}. Previous photochemical models have assumed a constant homopause pressure level with latitude, effectively imposing the same K$_{zz}$ profile at all latitudes, and therefore have not been able to capture this behavior.

Recent observations of Saturn's northern hemisphere by JWST revealed a depletion of CH\textsubscript{3} at high latitudes near the polar vortex \citep{Fletcher23}. \citet{Fletcher23} suggested that the low abundances of CH\textsubscript{3} at high latitudes could be attributed to downwelling in the upper stratosphere and lower thermosphere, resulting in the depression of the methane homopause to deeper pressures. This discovery establishes a possible link between our findings regarding the homopause level and novel findings in the infrared from JWST. Our observations of the methane homopause level also align well with Voyager 1 UVS observations analyzed by \citet{VM15}, while the Voyager 2 UVS results from that study indicate a somewhat lower-pressure homopause (Figure \ref{fig:Homo Trend}). The Voyager observations were made in early northern spring, with the Voyager 1 observations made when the sub-solar latitude was at 3.9$^{\circ}$N and the Voyager 2 occultations approximately 285 Earth days later when the sub-solar latitude was 8.1$^{\circ}$N. \citet{VM15} suggested that the decrease in the observed homopause pressure between the Voyager flybys could have a seasonal origin but it is not clear if the timescale between the flybys is sufficient to facilitate such a substantial change in the pressure level of the homopause. In any case, the homopause pressure levels retrieved from the Voyager observations show a similar meridional trend to our present results, with the lowest pressure homopause observed near the subsolar point.

We find that the base of the thermosphere coincides with the homopause. This is clear in Figure \ref{fig:Temp}, which shows the 2D distribution of temperature in the stratosphere, mesosphere, and thermosphere based on our results. In this figure, we also show the pressure level of the homopause at each occultation latitude. The base of the thermosphere is associated with a corresponding shift in composition. Methane and the photochemical products are largely confined to pressures greater than the homopause that coincide with temperatures less than about 225 K and in most cases below the region of the thermospheric temperature gradient. While the absorption of near-IR solar light by CH$_4$ bands heat the stratosphere, CH$_4$ and the hydrocarbons C$_2$H$_6$ and C$_2$H$_2$ are important radiative coolants in the mesosphere \citep{Yelle01}. Since the abundance of CH$_4$ falls rapidly with altitude above the homopause, the lack of radiative cooling in the thermosphere could explain the coexistence of the homopause and the base of the thermosphere, as it does on the Earth.

On the other hand, because of the influence of strong temperature gradients on wave propagation, it is possible that the steeply rising temperatures at the base of the thermosphere and associated steep increase in the molecular diffusion coefficients could give rise to the observed shape of the homopause level. Vertically propagating waves tend to break or dissipate in regions where molecular diffusion and thermal conduction increase rapidly with altitude \citep[e.g.,][]{medvedev19}. If the meridional distribution of Joule heating, circulation, and other factors in the thermosphere drive the base of the thermosphere to deeper pressure levels at the poles, that could confine wave breaking at those latitudes to deeper {pressure levels and inhibit vertical mixing of CH$_4$ and other hydrocarbons to the thermosphere. This would mean that along the isobars above the polar base of the thermosphere, we would expect greater vertical eddy mixing at low latitudes relative to the high latitudes, which could be contributing to or causing the larger values of K$_{zz}$ that are necessary for us to fit the data at low latitudes. In an apparent contrast to this idea, \citet{Brown22} characterized gravity waves based on the temperature and density profiles retrieved from the Grand Finale occultations. While wave activity was not sampled near the equator, the results indicated that wave breaking is occurring at middle and high latitudes above the homopause. However, this doesn't necessarily negate the hypothesis of weaker mixing driven by the base of the thermosphere at high latitudes because it is known that some waves can propagate past such temperature gradients to higher altitudes \citep[e.g.,][]{medvedev19}.

\begin{figure}
\noindent\includegraphics[width=\textwidth]{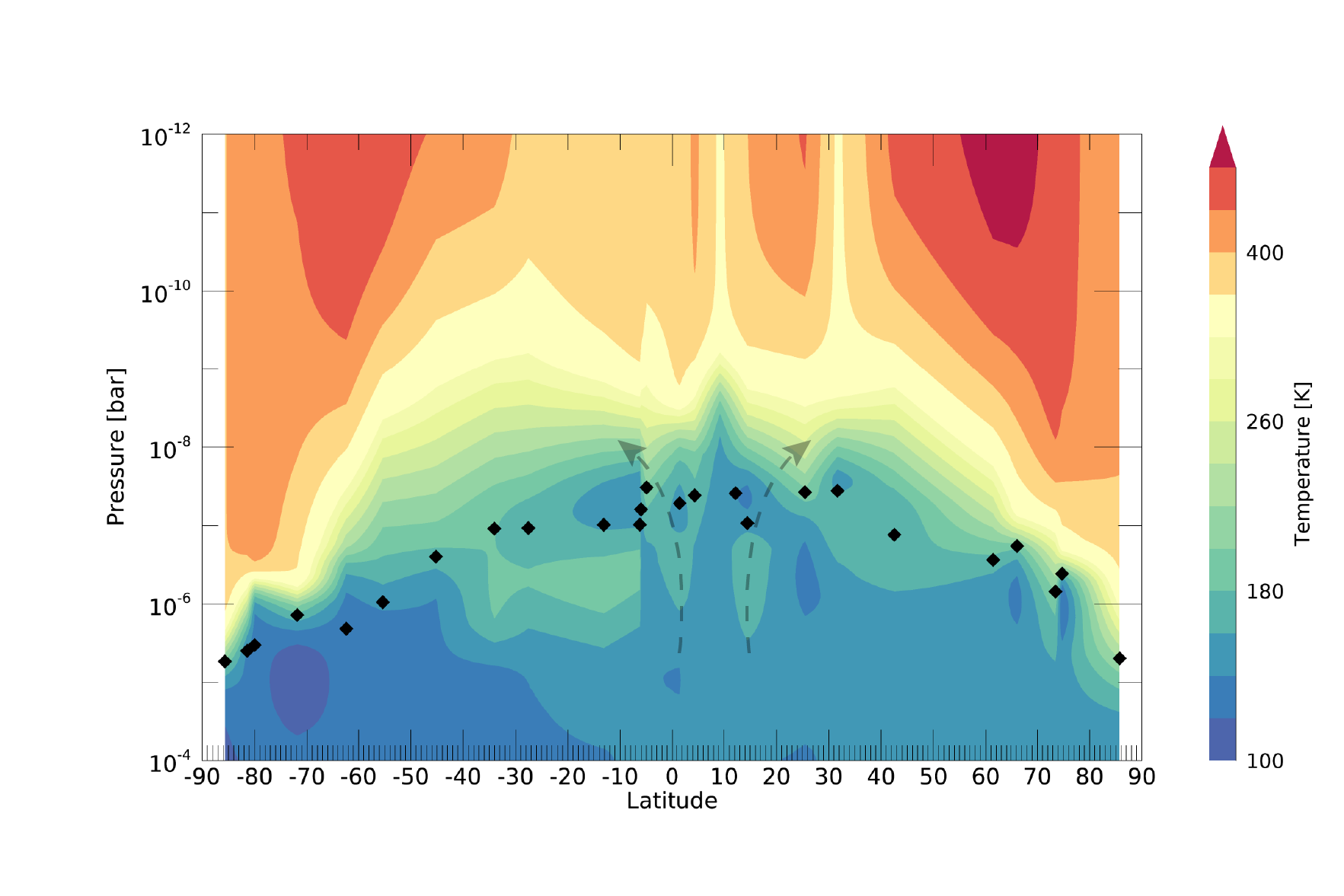}
\caption{Contour plot of temperatures from the atmospheric structure model, using CIRS observations from 2015 and 2017 at pressures higher than 10$^{-6}$ bar to 3~$\times$~10$^{-4}$ bar and UVIS observations at pressures lower than 10$^{-6}$ to 2~$\times$~ 10$^{-8}$ bar, where the quoted pressure limits depend on latitude. The temperatures in the coverage gap between the instruments were filled in to match the H$_2$ density vs. radius in the thermosphere. The black diamonds show the location of the CH$_4$ homopause, taken where CH$_4$ volume mixing ratio equals 5~$\times$~10$^{-5}$. A cartoon illustrating speculated low-latitude upwelling is depicted with dashed arrows near the high-altitude homopause levels.}
\label{fig:Temp}
\end{figure}

The eddy diffusion coefficient K$_{zz}$ profile that we use to match the observations simply parameterizes the extent of vertical mixing in the atmosphere and is agnostic to the source of vertical mixing. The alternative to the above scenario of weaker mixing induced by the base of the thermosphere is that the trend that we observe is caused by advection and in particular, vertical winds associated with global circulation. Indeed, vertical transport is considered the primary factor in determining observed abundances at a given location and is typically assumed to be the dominant factor in 1D models. However, meridional advection plays a crucial role in supporting vertical advection by sustaining the large-scale convection cells and can also play a role in controlling the meridional distribution of tracers \citep[For a detailed discussion of the theoretical background of tracer transport and circumstances in which advection produces different and similar outcomes to eddy diffusion, see][]{ZS18}. This would mean that relative upwelling takes place near the subsolar region and relative downwelling occurs at high latitudes that is enhanced in the winter hemisphere. This is currently our preferred explanation for two reasons. First, we required a similar variation in the $K_{zz}$ values in the stratosphere at 1 mbar level to match the observed C$_2$H$_2$ and C$_2$H$_6$ distributions (see Section~\ref{M-JModel}). It is hard to see how weaker mixing would persist over several decades in pressures at middle to high latitudes unless it is effectively due to downwelling associated with circulation. Second, observations of the stratosphere point to the existence of seasonal meridional circulation in the stratosphere that is also predicted by models (see Section~\ref{sc:intro}). This type of a circulation cell, however, would be limited to pressures greater than about 0.01 {$\mu$bar}. At lower pressures, equatorward winds are needed to redistribute energy from the poles to the equator and explain the observed temperatures throughout the thermosphere. 

In order to crudely estimate the relative magnitude of the velocities that would be required to match the observations by invoking vertical winds, we assume a constant K$_{zz}$ with latitude and use a simple timescale argument to constrain the relative downwelling and upwelling speeds. For example, we consider the 2 $\mu$bar level (where we have coverage of CH$_4$ at all sampled latitudes) and set K$_{zz}$ = 100 m$^2$~s$^{-1}$ for all latitudes. This value falls roughly half-way between the maximum and minimum values of K$_{zz}$ in the latitude space at this pressure level. The timescale for eddy diffusion is $H^2$/K$_{zz}$ and the vertical transport timescale is H/$v_z$, where H is the atmospheric scale height and $v_z$ is the vertical wind speed. Thus, we can crudely estimate the required vertical wind speed from:

\begin{equation}
v_z \approx (K_{zz}^{eff} - K_{zz}^{const})/H
\label{eqn:timescale}
\end{equation}
where $K_{zz}^{eff}$ is the value of K$_{zz}$ fitted to the data in our atmospheric structure models and $K_{zz}^{const} =$~100 m$^2$~s$^{-1}$. Figure \ref{fig:Winds} shows the resulting estimates of the vertical wind velocity as a function of latitude. In this case, positive upward velocities are required between 45$^{\circ}$ S and 60$^{\circ}$ N while downwelling occurs at higher latitudes. We note that the boundaries between upwelling and downwelling regions in our calculation are entirely dependent on the choice of $K_{zz}^{const}$. Furthermore, the upwelling at low latitudes and pressures greater than approximately 0.1 $\mu$bar should be regarded with caution since the K$_{zz}$ is poorly constrained below the homopause level. A coupled model of meridional circulation and photochemistry, fitted to the CH$_4$ and minor hydrocarbon profiles could be used to constrain the required K$_{zz}$ coefficient values and wind pattern simultaneously because vertical winds have a distinct signature on the observed profiles \citep{Moses15}. This is beyond the scope of the current work and will be pursued in future work.

The vertical wind speeds that we infer have an average magnitude of about 3 mm~s$^{-1}$, with a range from -4 mm~s$^{-1}$ to 17 mm~s$^{-1}$. These wind speeds are faster than previous analyses of Saturn's stratosphere have inferred from stratospheric temperature profiles observed by CIRS in the nadir mode (0.1 mm~s$^{-1}$) \citep{Flasar05}, the distribution of light hydrocarbons in the stratosphere from CIRS limb scans (0.25 mm~s$^{-1}$) \citep{Guerlet09}, and from polar C$_2$H$_2$ and C$_2$H$_6$ abundances observed in the stratosphere (a fraction of a mm~s$^{-1}$) \citep{FLETCHER15}. However, our derived maximum wind speeds are of the same order as the vertical speeds in Saturn's northern storm beacon region \citep{Moses15}. In addition, the winds speeds inferred from observations of the stratosphere referenced above apply at deeper pressure levels and there are no other observational constraints on the mesospheric wind speeds than our results here.

The interpretation of the homopause trend that relies on global circulation suggests the possibility of two circulation cells in the upper stratosphere and mesosphere, one in each hemisphere, characterized by upwelling near the subsolar point, meridional transport towards the poles, and downwelling at high latitudes that is strongest in the winter hemisphere. The stratospheric circulation model of \citet{Bardet22} predicts that this type of circulation should occur in the stratosphere near solstice while equatorial upwelling with downwelling at middle latitudes occurs at equinox. Our findings suggest that the predicted circulation may extend to lower pressures, reaching up to the 0.01 $\mu$bar pressure level. The distribution of upwelling and downwelling zones that we observe could be seen as transitional between these two patterns given that there could be a time lag in the response of the atmosphere to seasonal forcing. As we note above, the presence of such seasonally reversing meridional circulation in the stratosphere is also supported by several other observations \citep[e.g.,][]{Fletcher20}.

\begin{figure}
\noindent\includegraphics[width=\textwidth]{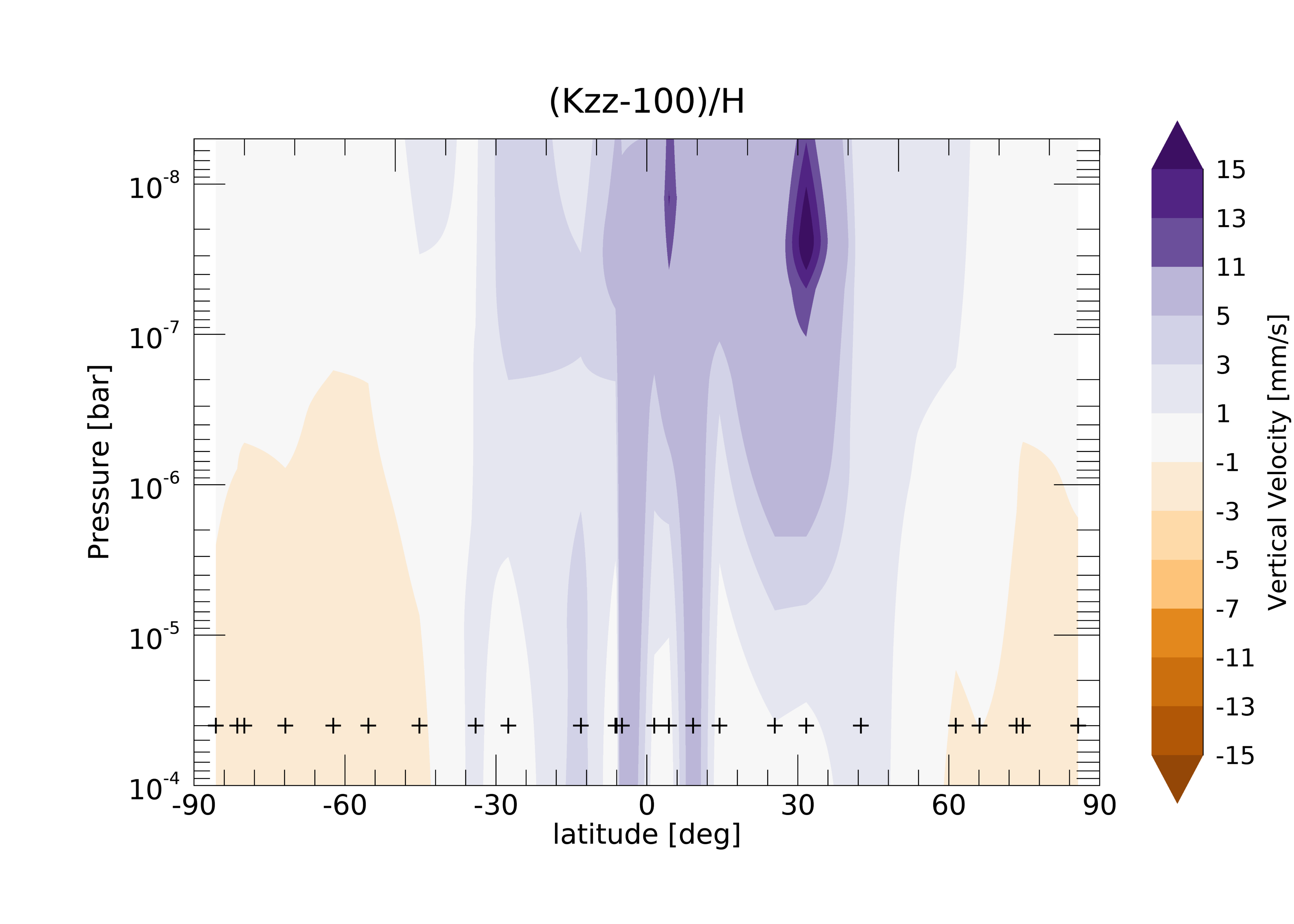}
\caption{Contour plot of vertical wind speeds estimated from equation~(\ref{eqn:timescale}). Upward velocities peak at 17 mm~s$^{-1}$ near 30$^{\circ}$ while downward velocities occur at higher latitudes, peaking at 4 mm~s$^{-1}$ in the southern hemisphere. The plus signs near the bottom of the plot indicate occultation latitudes.}
\label{fig:Winds}
\end{figure}

We conclude that upwelling in the summer hemisphere and downwelling in the winter hemisphere is a plausible explanation of the homopause trend that we observe in the Cassini/UVIS data. A secondary circulation cell in the northern hemisphere and associated downwelling at high latitudes could then explain the deeper homopause at northern summer latitudes. This would mean that the distribution of the hydrocarbons is also influenced by seasonally changing dynamics, in addition to changes in insolation. While the concerted EUV and FUV datasets observed by Cassini before and during the Grand Finale mission make this detailed view of the mesosphere and thermosphere possible at northern summer solstice, comparisons to other seasons would provide valuable insights. To investigate the annual behavior of Saturn's mesosphere, a model of meridional transport consistent with the predicted circulation could be developed and then applied to other seasons. The predictions of the model could be compared to, for example, Cassini dayglow observations \citep{Shemansky09,Koskinen20} to monitor seasonal variations and their effect on the depth of the homopause.

Interestingly, recent work by \citet{Sinclair20} presented contrasting findings regarding the homopause on Jupiter that could point to significant differences between the two planets. Using IRTF/TEXES, they observed CH$_4$, CH\textsubscript{3} and H\textsubscript{2} S(1) quadrupole emissions from the northern and southern hemispheres at latitudes ranging from 44$^{\circ}$ to just below 90$^{\circ}$. They used a family of photochemical model profiles with different K$_{zz}$ profiles to constrain the CH\textsubscript{3}/CH$_4$ ratio and found that the data are best fit with a homopause that lies at higher altitudes inside the auroral oval than at other latitudes. In these retrievals, however, temperature and abundances are degenerate and a higher CH\textsubscript{3} abundance, for example due to auroral chemistry, could fit the observations with a lower homopause altitude. In contrast to their results, \citet{Kim17} analyzed the north polar bright spot (NPBS) of Jupiter by fitting CH$_4$ emission models to 8 micron spectra observed with IRTF/TEXES and Gemini/GNIRS. Using a radiative transfer model, they compared the observations with mixing ratio profiles from the photochemical model of \citet{Moses05} and the temperature-pressure profile retrieved from the Galileo probe data by \citet{Seiff98}. By assessing the goodness of fit across multiple iterations of the \citet{MG05} model featuring different homopause levels, they found that the homopause altitude within the NPBS was similar to the one retrieved with similar methods by \citet{Kim14} for equatorial regions. Ultraviolet occultations by Jupiter's middle and upper atmosphere at different latitudes, expected for example from the Jupiter Icy Moons Explorer (JUICE) mission, would help to resolve the differences between these different interpretations.

\subsection{Seasonal Behavior of Hydrocarbons}
\label{D-Hydro}

\noindent
In this study, we retrieved hydrocarbon abundances, including C$_2$H$_6$, C$_2$H$_2$, C$_2$H$_4$, and C$_6$H$_6$, in the upper stratosphere and mesosphere, below and near the homopause. The most obvious trend in the minor hydrocarbon abundances is driven by the meridional trend in the homopause depth. Photochemistry is initiated by the dissociation of CH$_4$ by solar UV radiation, and subsequent hydrocarbon chemistry is constrained to pressure levels below the homopause. The trend observed in the retrieved occultation data reveals a rapid decrease in the concentrations of minor hydrocarbons with latitude away from the equator along isobars in the mesosphere at pressures lower than a few $\mu$bar. 

Beyond the trend in the depth of the homopause, our seasonal photochemical model predicts enhanced abundances of C$_2$H$_6$, C$_2$H$_2$, and C$_2$H$_4$ in the summer hemisphere due to the higher solar insolation and relatively short chemical timescale in the upper stratosphere and mesosphere. This trend is clearly seen in our observed hydrocarbon distributions, confirming the expectation that the abundances of photochemical products in the upper stratosphere and mesosphere vary strongly with season. The agreement between the observed hydrocarbon abundance profiles and those predicted by our photochemical model is relatively good at low and middle latitudes in the summer hemisphere, although the models tend to slightly overestimate the peak abundance of C$_2$H$_4$ (this is the case at all latitudes except those poleward of 65$^{\circ}$ N). The abundances of C$_2$H$_6$, C$_2$H$_2$, and C$_2$H$_4$ predicted by the Ion Model tend to be somewhat lower than those predicted by the Neutral Model, but the comparison with the observations remains qualitatively similar for these species.

The most obvious indication of the importance of ion chemistry in the observations are the retrieved C$_6$H$_6$ profiles. In the summer hemisphere, the C$_6$H$_6$ abundance predicted by the Ion Model is about an order of magnitude higher than the abundance predicted by the Neutral Model. As such, it provides a much better fit to the observations at northern low and middle latitudes than the neutral chemistry model. In the southern (winter) hemisphere at low and middle latitudes where the photon flux is lower, the model predicts an abundance too low to be detected in the occultations and indeed, we do not detect C$_6$H$_6$ in any occultation at latitudes between 20$^{\circ}$S and 50$^{\circ}$S. At the same time, the ion chemistry in the model is clearly not complete because the model underpredicts the observed peak C$_6$H$_6$/CH$_4$ volume mixing ratio at most latitudes, although simplifications in the C$_6$H$_6$ molecular diffusion coefficient assumptions also likely play a role \citep[see][and discussion above in Section \ref{M-JModel}]{MOSES23}. In general, the C$_6$H$_6$ abundances peak in the summer hemisphere (with local maxima near auroral latitudes in each hemisphere), exhibiting a strong seasonal trend, along with the other hydrocarbons.

In the winter hemisphere, the agreement between our models and the observed profiles is notably worse than in the summer hemisphere. At low latitudes, the observed peak abundances of C$_2$H$_6$, C$_2$H$_2$, and C$_2$H$_4$ are higher than the model abundances. However, the observed profiles still exhibit a peak indicative of in situ production by photochemistry. The presence of the ring shadow appears to impact the low-latitude winter models while not exerting a discernible influence on the corresponding observational data. At higher latitudes in the ring shadow and permanent night, even the shape of the observed profiles differs qualitatively from the photochemical model predictions. This is not surprising given the lower photon flux in the winter hemisphere where meridional and vertical transport by circulation that are not included in our photochemical model are expected to play a more important role. In the ring shadow region (between 17$^{\circ}$S to 31.20$^{\circ}$S, depending on the time of day, and extending into the permanent night at northern summer solstice), the functional form of the K$_{zz}$ profile imposed in our atmospheric structure model (equation 4 in \citealt{Yelle08}) struggles to reproduce the CH$_4$ profile and this enhances the discrepancy between the models and the observations for other species. The observed profiles of C$_2$H$_2$ and C$_2$H$_4$ appear flatter with pressure than predicted by the models and in contrast to the summer hemisphere, the abundance of C$_2$H$_2$ decreases with altitude between 100 and 1 $\mu$bar. These characteristics could arise from vertical transport associated with global circulation, although the feasibility of this explanation would need to be verified by coupled models of circulation and photochemistry. 

\subsection{Chemistry at Auroral Latitudes}
\label{D-Auroral}

\noindent
At high latitudes near and around the auroral ovals in the summer hemisphere, there are more significant differences between our photochemical models and the observations. These differences are likely due, at least in part, to the lack of production initiated by auroral electron and ion precipitations in our models, although they could also be related to missing transport processes. At low latitudes, the observed abundance of C$_2$H$_6$ is clearly higher than the abundance of C$_2$H$_2$, however, at higher northern latitudes the abundance of C$_2$H$_2$ begins to catch up to the C$_2$H$_6$ abundance; between 0.3 and 0.5 $\mu$bar, C$_2$H$_6$ is four times more abundant than C$_2$H$_2$ at 31.7$^\circ$N, whereas at 61.4$^\circ$N and 74.7$^\circ$N, C$_2$H$_6$ is about twice as abundant as C$_2$H$_2$. This behavior disagrees with the model predictions for the relative abundances of these species. This is demonstrated in Figure \ref{fig:meridionaltrends}, where the northernmost C$_2$H$_6$ point at 85.7$^{\circ}$N is approximately equal to the northernmost observable C$_2$H$_2$ point at 42.5$^{\circ}$N. Thus, the observed abundance of C$_2$H$_2$ around the northern auroral oval is significantly larger than the abundances predicted by the model, and the inclusion of solar-driven ion chemistry does not help to resolve this discrepancy. Retrievals  of C$_2$H$_2$ and C$_2$H$_6$ on Jupiter from mid-infrared data indicate that C$_2$H$_2$ is enriched in auroral regions relative to quiescent regions while C$_2$H$_6$ is depleted \citep{Sinclair17}. They point to the role ion-neutral and electron recombination chemistry in the auroral region in forming unsaturated hydrocarbons like C$_2$H$_2$, which may explain the increased abundance of this molecule present in our high latitude observations. To explain their observed distribution of C$_2$H$_6$, they note that if unsaturated hydrocarbons are transported outside of the auroral region by winds, they could then be converted to C$_2$H$_6$ by neutral photochemistry, enriching the C$_2$H$_6$ abundance observed outside of the auroral ovals. Our results on Saturn, however, indicate that the relative abundances of C$_2$H$_2$ and C$_2$H$_6$ begin to change at latitudes higher than about 60$^{\circ}$N, in a region well outside of the auroral ovals. The solution to this problem may therefore be more complicated than the simple inclusion of auroral chemistry or ion chemistry in the model. Given the deeper homopause at high latitudes, mixing and transport could also play a role in shaping the observed profiles.

The observed C$_6$H$_6$ abundances provide the clearest evidence available for additional production at high latitudes near the auroral ovals. In the northern auroral region, the observed C$_6$H$_6$ abundances exceed our model predictions by 1 to 2 orders of magnitude. The discrepancy in the southern hemisphere is even worse, with the observed C$_6$H$_6$ abundance exceeding our model predictions by more than two orders of magnitude in the region of permanent night. Some caution, however, is required to interpret this result. C$_6$H$_6$ appears in the southern hemisphere at latitudes higher than about 55$^{\circ}$S, well outside of the auroral oval, indicating that transport also plays a role in controlling the distribution of hydrocarbons in the winter hemisphere. In addition, it is not clear what fraction of auroral electrons or ions are sufficiently energetic to penetrate past the deep homopause. Naively, one might assume that their impact is limited below the base of the thermosphere (assuming that they also heat the auroral thermosphere), in which case the direct impact of auroral particle precipitation could be more limited than anticipated. Our results clearly call for new models of Saturn’s middle and upper atmosphere that include auroral chemistry and transport by atmospheric circulation. These results provide some of the best constraints on such models available for any giant planet, enhancing the already remarkable legacy of the Cassini mission.

\section{Summary and Conclusions}

\noindent
This study provides the first 2D maps of the CH$_4$, C$_2$H$_6$, C$_2$H$_2$, C$_2$H$_4$, and C$_6$H$_6$ distributions in Saturn's upper stratosphere and mesosphere for a fixed season retrieved from stellar occultations observed by the Cassini UVIS instrument. Temperature profiles retrieved from the same occultations and CIRS limb scans, along with the abundances of C$_2$H$_6$ and C$_2$H$_2$ retrieved from the CIRS data, are used to inform atmosphere structure models that permit for the observed densities to be converted to volume mixing ratios as a function of pressure. This allows for the comparison of the observations with photochemical model predictions over a wide range of pressures. A key finding of this study is a pronounced meridional variation in the observed homopause pressure level, with a much deeper homopause around the poles than around the subsolar point. This is observable as a large and clear variation in CH$_4$ mixing ratio along isobars at pressures lower than about 10 $\mu$bar and it necessitates the introduction of a latitude-dependent K$_{zz}$ profile to accurately model the CH$_4$ abundances. These results indicate weaker mixing at high latitudes than near the subsolar point at pressures between 10 and 0.01 $\mu$bar. The homopause reference level coincides closely with the base of thermosphere i.e., the pressure where temperature begins to rise rapidly with altitude. The base of the thermosphere is deeper around the poles and it is possible that dynamics and Joule heating in the deeper thermosphere drives the observed hydrocarbon distributions by inhibiting the mixing of CH$_4$ up along the steep temperature gradient. 

Our preferred explanation for the homopause trend, however, is a global circulation pattern consisting of upwelling near the subsolar latitude and downwelling at high latitudes that is more robust in the winter hemisphere. We prefer this interpretation because of the similarity of the $K_{zz}$ profiles inferred from the UVIS data to the ones that are needed to match the hydrocarbon distributions deeper in the stratosphere where seasonally reversing meridional circulation is predicted by models and supported by observations. Using a simple timescale argument, we find that an average vertical wind speed of 3 mm~s$^{-1}$, with a range from -4 mm~s$^{-1}$ to 17 mm~s$^{-1}$ would be required to reproduce the observed trend. These vertical wind speeds are larger than the values inferred from observations of the stratosphere (0.1 mm~s$^{-1}$ \cite{Flasar05}, 0.25 mm~s$^{-1}$ \cite{Guerlet09}, 1 mm~s$^{-1}$ \cite{FM12}, fractions of a mm~s$^{-1}$ \cite{FLETCHER15}), but consistent with the higher wind speeds expected at lower pressures in the upper stratosphere and mesosphere of Earth \citep[to 10s of m~s$^{-1}$,][]{LM12}, \citep[2 cm~s$^{-1}$,][]{Swenson19}.

Consistent with the predictions of the photochemical models, we detect a clear seasonal trend in the observed abundances of the photochemical products at pressures lower than about 0.01 mbar, with higher abundances overall in the summer hemisphere than in the winter hemisphere. One exception is the abundance of C$_2$H$_2$ near 20$^\circ$S latitude, which exceeds model predictions. As in \citet{Guerlet09}, the existence of increased levels of this short-lived molecule in the winter hemisphere at 0.01 mbar, could point to summer to winter transport. The greater seasonal trend observed at low pressures differs from deeper pressure levels in the stratosphere where photochemical timescales are long enough to suppress seasonal trends in the abundances of C$_2$H$_2$, C$_2$H$_6$, and other long-lived hydrocarbons. To our knowledge, this is the first time that a clear seasonal trend has been detected, due to the more comprehensive coverage of the photochemical production region provided by the UVIS instrument during a fixed season. In general for this pressure range, photochemical models that do not include mixing by circulation agree better with the observed abundance profiles in the summer hemisphere than in the winter hemisphere. This is expected because of the higher solar insolation in the summer hemisphere that drives the profiles closer to seasonal photochemical equilibrium. Dynamics is expected to play a more important role in shaping the abundance profiles in the ring-shadowed winter hemisphere, and this expectation is borne out by the UVIS observations. 

Out of the species that we detect in the UVIS observations, C$_6$H$_6$ is the clearest tracer of ion chemistry. Photochemical models that only include neutral chemistry underpredict its abundance by several orders of magnitude. \citet{Koskinen16} used simple calculations based on a few of the key reaction rates to argue that ion chemistry would improve the fit to the previously available occultation results on C$_6$H$_6$. Here, we incorporated the solar-driven ion chemistry explicitly in our photochemical model, and more than doubled the number of reactions that it includes as a result. The new model leads to a much better agreement with the observations, although the predicted abundances still remain somewhat lower than the observed abundances at most latitudes. The differences between the ion chemistry model and the observed C$_6$H$_6$ are particularly notable at high latitudes in the winter hemisphere and at high latitudes near the northern auroral ovals where auroral electron and ion precipitation may contribute to the production rates. The agreement between the model and the observations is most reasonable at northern low and middle latitudes. At southern middle latitudes in the ring shadow region, the model predicts low abundances of C$_6$H$_6$ that are not detectable in the occultations and this expectation is borne out by the observations that show no evidence for absorption by C$_6$H$_6$ at these latitudes. 

In the northern auroral oval we also observe a pronounced increase in the abundance of C$_2$H$_2$ over values predicted by our photochemical model in the production region between 1 and 10 $\mu$bar. The observed C$_2$H$_2$/C$_2$H$_6$ abundance ratio also increases with latitude towards the poles, while the abundances of C$_2$H$_4$ and C$_2$H$_6$ (see Figures \ref{fig:VMR} and \ref{fig:ndensprofile}) are also generally higher than expected at these latitudes and pressures. These discrepancies are not resolved by the inclusion of ion chemistry. Our results therefore indicate that auroral chemistry may play an important role in both hemispheres. At the same time, it seems clear that transport is also affecting the observed profiles, particularly in the winter hemisphere. Taken together, these findings emphasize the need for comprehensive models that incorporate both auroral chemistry and atmospheric circulation to provide a more complete understanding of Saturn's middle and upper atmosphere dynamics. The observations described in this work provide critical constraints for such models, contributing to our growing knowledge of middle and upper atmosphere processes on this intriguing giant planet as well as enriching the legacy of the Cassini mission.

\section*{Acknowledgements}
\noindent
ZLB, TTK, and JM were supported by the NASA Cassini Data Analysis Program grant 80NSSC19K0902. JM acknowledges some additional support for the neutral seasonal models from NASA Solar System Workings Program 80NSSC20K0462. We would like to thank the reviewers for their thoughtful comments and efforts towards improving our manuscript.

\section*{Declaration of Use of Generative AI}
\noindent
During the preparation of this work the authors used ChatGPT in order to improve readability and language. After using this tool/service, the authors reviewed and edited the content as needed and take full responsibility for the content of the publication.

\section*{Data Availability}
\noindent
Hydrocarbon number density data observed with UVIS stellar occultations as well as the pressures, temperatures, atmospheric densities and K$_{zz}$ profiles derived from the atmospheric structure model are available in Mendeley Data, V1, https://doi.org/10.17632/67f3y7vb85.1. The unprocessed stellar occultation spectral cubes are available from the NASA Planetary Data System, https:// pds-imaging.jpl.nasa.gov/portal/cassini\textunderscore mission.html. CIRS hydrocarbon abundances and temperatures are available in Mendeley Data, V1, doi: 10.17632/nmmxfvhgh4.1. Photochemical model results for the Ion and Neutral Model for L\textunderscore s = 90, along with full seasonal model results for the year 6 of the Neutral model at 10 degree increments of L\textunderscore s are available in Mendeley Data, V2, https://doi: 10.17632/hhx6hytkj2.2.


\appendix

\section*{Appendix A. The Functional Form of K$_{zz}$ in the Photochemical Model}
\label{sec:Appendix_Kzz}

In our photochemical model, the eddy diffusion coefficient K$_{zz}$, is parameterized as a function of pressure and latitude using the following equations:

\renewcommand{\theequation}{A.\arabic{equation}}
\begin{equation}
    K_{zz}(p,\lambda) = K_0(\lambda) \left(\frac{p_0(\lambda)}{p(\lambda)}\right)^{\alpha_0} p < p_0
    \label{eq:JMKzz}
\end{equation}

\begin{equation}
    K_{zz}(p,\lambda) = K_1(\lambda) \left(\frac{p_1(\lambda)}{p(\lambda)}\right)^{\alpha_1}, p_0 < p < p_1
    \label{eq:JMKzz1}
\end{equation}

\begin{equation}
    K_{zz}(p,\lambda) = K_2(\lambda) \left(\frac{p_2(\lambda)}{p(\lambda)}\right)^{\alpha_2}, p_1 < p < p_2
    \label{eq:JMKzz2}
\end{equation}

\begin{equation}
    K_{zz}(p,\lambda) = K_3(\lambda) \left(\frac{p_3(\lambda)}{p(\lambda)}\right)^{\alpha_3}, p_2 < p < p_3
    \label{eq:JMKzz3}
\end{equation}

\noindent Here, K$_0$ through K$_3$ and p$_0$ through p$_3$ represent reference values of K${zz}$ and pressure, respectively, with both varying with latitude, $\lambda$, and applicable in subsequent pressure regimes. These K$_n$ and p$_n$ values, along with the parameter $\alpha$, are adjustable constants in each fit. They are selected to align with the C$_2$H$_6$ profile derived from CIRS in the lower stratosphere and the CH$_4$ profile from UVIS in the upper stratosphere. In most cases, K$_{zz}$ is set to a constant value above a defined pressure level. In the troposphere, it is set to another constant value to account for potentially increased mixing from convection. The functional form of K$_{zz}$ along with the reference values p$_0$ and K$_0$ may vary with latitude to better fit observations from CIRS and UVIS.


\bibliographystyle{elsarticle-harv}
\bibliography{refs.bib}





\end{document}